\documentclass[preprint]{llncs}
\pdfoutput=1
\usepackage[usenames]{color}
\usepackage{graphics}
\usepackage{url}
\usepackage{qsymbols}
\usepackage{amsmath}
\usepackage{amssymb}
\usepackage{pifont}
\usepackage{stmaryrd}
\usepackage{subfig}
\usepackage{xspace}
\usepackage{mathtools}
\usepackage{booktabs}

\usepackage{centernot}

\usepackage[T1]{fontenc} 
\usepackage{lmodern} 
\IfFileExists{luximono.sty}{\usepackage[scaled=0.9]{luximono}}{\usepackage[scaled=0.81]{beramono}}

\usepackage{silver}

\usepackage{etoolbox}
\expandafter\patchcmd\csname \string\lstinline\endcsname{%
  \leavevmode
  \bgroup
}{%
  \leavevmode
  \ifmmode\hbox\fi
  \bgroup
}{}{%
  \typeout{Patching of \string\lstinline\space failed!}%
}

\usepackage{xcolor}
\usepackage{wrapfig}

\DeclarePairedDelimiter{\parens}{(}{)}

\newcommand*{\thmref}[1]{Thm.\,\ref{thm:#1}}
\newcommand*{\lemref}[1]{Lemma\,\ref{lem:#1}}
\newcommand*{\defref}[1]{Def.\,\ref{def:#1}}
\newcommand*{\figref}[1]{Fig.\,\ref{fig:#1}}
\newcommand*{\secref}[1]{Sec.\,\ref{sec:#1}}
\newcommand*{\appref}[1]{{\color{red}App.\,? in \trcite}}
\newcommand*{\appsecref}[1]{\secref{#1}}
\newcommand*{\appfigref}[1]{\figref{#1}}
\newcommand*{\tabref}[1]{Table\,\ref{tab:#1}}
\newcommand{\trcite}{{\color{red}\cite{?}}}

\makeatletter
\newcommand{\accSym}{\textsf{acc}}
\newcommand{\acc}{\ensuremath{\@ifnextchar\bgroup {\accOne}{{\accSym}}}}
\newcommand{\accOne}[1]{\ensuremath{\@ifnextchar\bgroup {\accT{#1}}{{\accSym}(\texttt{#1})}}}
\newcommand{\accT}[2]{{\accSym}({#1},{#2})}

\newcommand{\multOpSym}{\ensuremath{\odot}}
\newcommand{\mult}{\ensuremath{\@ifnextchar\bgroup{\multT}{\multOpSym}}}
\newcommand{\multT}[2]{\ensuremath{\code{{#2}}}}

\makeatother

\newcommand{\reduces}{\longrightarrow}
\newcommand{\reducesOne}{\reduces}
\newcommand{\reducesStar}{\reducesOne^{\ast}}

\newcommand{\imp}{\Rightarrow}

\newcommand{\range}[2]{\ensuremath{{[}{#1},{#2}{]}}}

\newcommand{\cf}{{{cf.}}}
\newcommand{\eg}{{{e.g.}}}
\newcommand{\ie}{{{i.e.}}}

\newcommand{\defequals}{\ensuremath{{:}{:}{=}}}

\makeatletter
\def\operator#1{\ensuremath{\@ifnextchar\bgroup {\operatorarg{#1}}{#1}}}
\def\operatorarg#1#2{{#1}{(#2)}}

\def\spoperator#1{\ensuremath{\@ifnextchar\bgroup {\spoperatorarg{#1}}{#1}}}
\def\spoperatorarg#1#2{{#1}{\mathtt{~}#2}}

\def\inhale{\operator{\mathtt{inhale}}}
\def\exhale{\operator{\mathtt{exhale}}}

\def\FV{\operator{\textit{FV}}}

\def\length{\operator{\mathtt{len}}}
\def\lcm{\operator{\textit{lcm}}}

\def\Inv{\INV} 
\def\INV{\INVplus}
\def\Invplus{\INVplus} 
\def\INVplus{\operator{\ensuremath{\mathcal{I}^{+}}}}
\def\Invminus{\INVminus} 
\def\INVminus{\operator{\ensuremath{\mathcal{I}^{-}}}}

\def\Vec#1{\ensuremath{\overline{#1}}}
\def\vec{\Vec}

\newcommand{\simple}[1]{\@ifnextchar\bgroup{\simpleTwo{#1}}{\simpleTwo{#1}{\x}}}
\newcommand{\simpleTwo}[2]{#1}

\newcommand{\arr}{a}

\newcommand{\moduloSym}{\textrm{mod}}
\newcommand{\modulo}{\@ifnextchar\bgroup{\moduloTwo}{\moduloSym}}
\newcommand{\moduloTwo}[2]{\ensuremath{(#1 \moduloSym #2)}}
\newcommand{\dividesSym}{\mid}
\newcommand{\divides}{\@ifnextchar\bgroup{\dividesTwo}{\dividesSym}}
\newcommand{\dividesTwo}[2]{\ensuremath{#2 {\modSym} #1{=}0}}
\newcommand{\notdividesSym}{\centernot\mid}
\newcommand{\notdivides}{\@ifnextchar\bgroup{\notdividesTwo}{\notdividesSym}}
\newcommand{\notdividesTwo}[2]{\ensuremath{#2 {\modSym} #1 {\neq} 0}}
\newcommand{\modSym}{\ensuremath{{\scriptstyle \%}}}

\newcommand{\veeunder}[2]{\ensuremath{\displaystyle{\bigvee_{#1}{#2}}}}

\newcommand{\maxdefn}[2]{\ensuremath{\displaystyle{\max_{#1}{#2}}}}
\newcommand{\inlinemaxunder}[2]{\textstyle{\max_{#1}{(#2)}}}
\newcommand{\maxunder}[2]{\@ifnextchar\bgroup{\maxunderThree}{\maxdefn{#1}{#2}}}
\newcommand{\maxunderThree}[3]{\maxunder{#1 \mid #2}{#3}}

\newcommand{\by}{\ensuremath{{\cdot}}} 

\newcommand{\deref}[2]{\ensuremath{{#1}[#2]}}
\newcommand{\lookup}[3]{\ensuremath{{#1}(#2,#3)}}

\newcommand{\preSym}{\textit{pre}}

\newcommand{\wppSym}{\preSym}
\newcommand{\wpp}{\ensuremath{\@ifnextchar\bgroup{\wppOne}{\wppSym}}}
\newcommand{\wppOne}[1]{\ensuremath{\@ifnextchar\bgroup{\wppTwo{#1}}{\wppDone{\i}{\arr}{#1}}}}
\newcommand{\wppTwo}[2]{\ensuremath{\@ifnextchar\bgroup{\wppThree{#1}{#2}}{\wppDone{\i}{\arr}{#1,#2}}}}
\newcommand{\wppThree}[3]{\ensuremath{\@ifnextchar\bgroup{\wppFour{#1}{#2}{#3}}{\wppDone{\i}{#1}{#2,#3}}}}
\newcommand{\wppFour}[4]{\wppDone{#1}{#2}{#3,#4}}
\newcommand{\wppDone}[3]{\wppSym(#3)} 

\newcommand{\diffSym}{\operatorname{\Delta}}
\newcommand{\diff}{\ensuremath{\@ifnextchar\bgroup{\diffOne}{\diffSym}}}
\newcommand{\diffOne}[1]{\ensuremath{\@ifnextchar\bgroup{\diffTwo{#1}}{\diffSym(#1)}}}
\newcommand{\diffTwo}[2]{\diffOne{#1,#2}}

\newcommand{\miniSym}{\textit{small}}
\newcommand{\mini}{\ensuremath{\@ifnextchar\bgroup{\miniOne}{\miniSym}}}
\newcommand{\miniOne}[1]{\ensuremath{\@ifnextchar\bgroup{\miniTwo{#1}}{\miniTwo{\x}{#1}}}}
\newcommand{\miniTwo}[2]{\ensuremath{\cornerBrackets{#2}_{\kern-2.5\point\miniSym(#1)}}}

\newcommand{\minimaxSym}{\textit{smallmax}}
\newcommand{\minimax}{\ensuremath{\@ifnextchar\bgroup{\minimaxOne}{\minimaxSym}}}
\newcommand{\minimaxOne}[1]{\ensuremath{\@ifnextchar\bgroup{\minimaxTwo{#1}}{\minimaxTwo{\x}{#1}}}}
\newcommand{\minimaxTwo}[2]{\ensuremath{\cornerBrackets{#2}_{\kern-2.5\point\minimaxSym(#1)}}}

\newcommand{\restrict}[1]{\ensuremath{\overapprox{#1}}}

\newcommand{\suchthat}{\ensuremath{\mid}}

\newcommand{\leftInfSym}{\ensuremath{-\infty}}
\newcommand{\leftInf}{\ensuremath{\@ifnextchar\bgroup{\leftInfOne}{\leftInfSym}}}
\newcommand{\leftInfOne}[1]{\ensuremath{\@ifnextchar\bgroup{\leftInfTwo{#1}}{\leftInfTwo{\x}{#1}}}}
\newcommand{\leftInfTwo}[2]{\ensuremath{\cornerBrackets{#2}_{\kern-2.5\point\leftInfSym(#1)}}}

\newcommand{\entails}{\ensuremath{\@ifnextchar\bgroup {\entailsargs}{\entailssymb}}}
\newcommand{\entailssymb}{\models}
\newcommand{\entailsargs}[2]{\ensuremath{#1 \entailssymb #2}}

\makeatother

\renewcommand{\a}{q_a}
\newcommand{\qi}{q_i}
\newcommand{\x}{x}

\def\trueval{\textup{\textsf{true}}}
\def\falseval{\textup{\textsf{false}}}
\def\trueexp{\textup{\textsf{true}}}
\newcommand{\falseExp}{\falseexp}
\newcommand{\trueExp}{\trueexp}
\def\falseexp{\textup{\textsf{false}}}

\newcommand{\permfail}{\ensuremath{\textup{\textsf{perm-fail}}}}

\newcommand\stackCloserHelp[2]{{%
    \setbox0\hbox{\ensuremath{#1}}%
    \rlap{\hbox to \wd0{\hss\raisebox{-.7\height}{#2}\hss}}\box0
}}

\makeatletter

\newcommand\union{\ensuremath{\cup}}

\newcommand{\op}{\ensuremath{~\textit{op}~}}

\def\op{\textit{op}}

\newcommand\noop{\ensuremath{\mathtt{skip}}}
\newcommand\seq[2]{\ensuremath{(#1;#2)}}
\newcommand\assign[2]{\ensuremath{#1{:}{=}#2}}
\newcommand\ifcond[3]{\ensuremath{\mathtt{if(} #1 \mathtt{)~\{~} #2 \mathtt{~\}~else~\{~} #3 \mathtt{~\}}}}
\newcommand\cond[3]{\ensuremath{(#1 \mathbin{?} #2 \mathbin{:} #3)}}
\newcommand\while[2]{\ensuremath{\mathtt{\whilesym~}(#1)\mathtt{~\{~} #2 \mathtt{~\}}}}
\newcommand\whilesym{\ensuremath{\mathtt{while}}}

\newcommand\upd[3]{\ensuremath{#1[#2{\mapsto}#3]}}

\newcommand\sub[3]{\ensuremath{#1[#2/#3]}}
\newcommand\vecsub[3]{\ensuremath{#1\vec{[#2/#3]}}}

\newcommand{\zeroPerm}{\ensuremath{0}}

\newcommand{\leaf}[2]{\cond{#1}{#2}{\zeroPerm}}
\newcommand{\accperm}[3]{\alpha_{#1,#2}\parens{#3}}

\newcommand\psub[4]{\ensuremath{#1[\deref{#2}{#3} \mapsto #4]}}

\newcommand\rd{\ensuremath{\textsf{rd}}}

\newcommand\state[1]{\@ifnextchar\bgroup{\statethree{#1}}{\ensuremath{(#1)}}}
\newcommand\statethree[3]{\state{#1,#2,#3}}

\newcommand\config[1]{\@ifnextchar\bgroup{\configtwo{#1}}{\ensuremath{{<}{#1}{>}}}}
\newcommand\configtwo[2]{\config{#1,#2}}

\def \evalExp#1{\@ifnextchar\bgroup{\evalExpTwo{#1}}{\sqBrackets{#1}{}}}
\def \evalExpTwo#1#2{\@ifnextchar\bgroup{\evalExpMore{#1}{#2}}{\sqBrackets{#1}{\state{#2}}}}
\def \evalExpMore#1#2#3{\evalExp{#1}{#2,#3}}

\newcommand{\overapprox}[1]{#1^\uparrow}
\newcommand{\underapprox}[1]{#1^\downarrow}
\newcommand{\overhavoc}[3]{#1^{\uparrow(#2,#3)}}
\newcommand{\underhavoc}[3]{#1^{\downarrow(#2,#3)}}

\newcommand{\more}{\ensuremath{\geq}}
\usepackage{prooftree}

\makeatletter

\newskip \point

\setbox136=\hbox{\leavevmode\raise0\point\hbox{${\lfloor}\kern-3\point{\lfloor}$}}
\setbox137=\hbox{\raise0\point\hbox{${ \rfloor}\kern-3\point{ \rfloor}$}}

\def \premisespacing{\quad}

\def \RulePremisesNewlineMore[#1]#2.#3#4{\@ifnextchar\bgroup{\RulePremisesNewlineMore[#1]{#2}.{#3\premisespacing#4}}{\@ifnextchar.{\RulePremisesNewline[#1]{{\begin{array}{c}#2\\#3\premisespacing#4\end{array}}}}{\RuleMultiPremise[#1]{{\begin{array}{c}#2\\#3\end{array}}}{#4}}}}

\def \RulePremisesNewline[#1]#2.#3{\@ifnextchar\bgroup{\RulePremisesNewlineMore[#1]{#2}.{#3}}{\@ifnextchar.{\RulePremisesNewline[#1]{{\begin{array}{c}#2\\#3\end{array}}}}{\RuleMultiPremise[#1]{#2}{#3}}}}

\def \RuleMultiPremise[#1]#2#3{\@ifnextchar\bgroup{\RuleMultiPremise[#1]{#2\premisespacing#3}}{\@ifnextchar.{\RulePremisesNewline[#1]{#2\premisespacing#3}}{\prooftree #2\justifies#3 \using{#1}\endprooftree}}}

\def \RuleWithName[#1]#2{\@ifnextchar\bgroup {\RuleMultiPremise[#1]{#2}}{\@ifnextchar.{\RulePremisesNewline[#1]{#2}}{\prooftree \justifies #2 \using{#1} \endprooftree}}}

\def \RuleWithInfo[#1]{\@ifnextchar[{\RuleWithNameAndCondition[#1]}{\RuleWithName[(\textit{#1})]}}

\def \RuleWithNameAndCondition[#1][#2]{\RuleWithName[(\textit{#1})^{#2}]}

\def \Inf{\proofrulebaseline=2ex \abovedisplayskip12\point\belowdisplayskip12\point \abovedisplayshortskip8\point\belowdisplayshortskip8\point \@ifnextchar[{\RuleWithInfo}{\RuleWithName[ ]}}

\makeatother

\setbox134=\hbox{\leavevmode\raise0\point\hbox{${\langle}\kern-2.5\point{\langle}$}}
\setbox135=\hbox{\raise0\point\hbox{${ \rangle}\kern-2.5\point{ \rangle}$}}
\def \cornerBrackets#1{\copy134{#1}\copy135}
\setbox136=\hbox{\leavevmode\raise0\point\hbox{${\lfloor}\kern-3.0\point{\lfloor}$}}
\setbox137=\hbox{\raise0\point\hbox{${ \rfloor}\kern-3.0\point{ \rfloor}$}}

\setbox138=\hbox{\leavevmode\raise0\point\hbox{${[}\kern-1.5\point{[}$}}
\setbox139=\hbox{\raise0\point\hbox{${]}\kern-1.5\point{]}$}}
\newcommand{\sqBrackets}[2]{\copy138{#1}\copy139_{#2}}
\setbox140=\hbox{\leavevmode\raise0\point\hbox{${\lfloor}$}}
\setbox141=\hbox{\raise0\point\hbox{${ \rfloor}$}}

\setbox142=\hbox{\leavevmode\raise0\point\hbox{${\lceil}$}}
\setbox143=\hbox{\raise0\point\hbox{${ \rceil}$}}

\usepackage[normalem]{ulem}

\begin{document}

\title{Permission Inference for Array Programs}

\author{J\'er\^ome Dohrau \and Alexander J.\ Summers \and \\ Caterina Urban \and Severin M\"unger \and Peter M\"uller}

\def\emailspacing{\ \ }
\institute{
Department of Computer Science, ETH Zurich, Switzerland\\
\email{jerome.dohrau@inf.ethz.ch}\emailspacing{}\email{severin.muenger@alumni.ethz.ch}\\\email{\{alexander.summers, caterina.urban, peter.mueller\}@inf.ethz.ch}
}

\maketitle
\pagestyle{plain}

\begin{abstract}
Information about the memory locations accessed by a program is, for
instance, required for program parallelisation and program
verification. Existing inference techniques for this information
provide only partial solutions for the important class of
array-manipulating programs. In this paper, we present a static
analysis that infers the memory footprint of an array program in terms of
permission pre- and postconditions as used, for example, in separation
logic. This formulation allows our analysis to handle concurrent
programs and produces specifications that can be used by verification
tools.
Our analysis expresses the permissions required by a loop via
maximum expressions over the individual loop iterations.
These maximum expressions are then solved by a novel maximum
elimination algorithm, in the spirit of quantifier elimination.
Our approach is sound and is implemented; an evaluation on
existing benchmarks for memory safety of array programs demonstrates
accurate results, even for programs with complex access patterns and
nested loops.
\end{abstract}

\section{Introduction}
\label{sec:introduction}

Information about the memory locations accessed by a program is
crucial for many applications such as static data race
detection \cite{Voung:2007}, code
optimisation \cite{JohnsonEtAl17,LernerGC02,FerranteOW84}, program
parallelisation \cite{GedellH06,BlomDH15}, and program
verification \cite{Rustan10,JacobsEtAl11,PiskacWiesZufferey14,MuellerSchwerhoffSummers16}. The
problem of inferring this information statically has been addressed by
a variety of static
analyses, e.g., \cite{CalcagnoDOY11,SalcianuR05}. However, prior works
provide only partial solutions for the important class of
array-manipulating programs for at least one of the following
reasons. (1)~They approximate the entire array as one single memory
location \cite{BertraneCCFMMR10}
which leads to imprecise results;
(2)~they do not produce specifications, which are useful
for several important applications such as human inspection, test case
generation, and especially deductive program verification;
(3)~they are limited to sequential programs.

In this paper, we present a novel analysis for array programs that
addresses these shortcomings. Our analysis employs the notion
of \emph{access permission} from separation logic and similar program
logics \cite{Reynolds02,Smans09}.  These logics associate a permission with each
memory location and enforce that a program part accesses a
location only if it holds the associated permission. In this setting,
determining the accessed locations means to infer a sufficient
precondition that specifies the permissions required by a program
part.

Phrasing the problem as one of permission inference allows us to address
the three problems mentioned above. (1)~We distinguish different array
elements by tracking the permission for each element separately.
(2)~Our analysis infers pre- and postconditions for both methods and
loops and emits them in a form that can be used by verification tools.
The inferred specifications can easily be complemented with permission
specifications for non-array data structures and with functional
specifications.
(3)~We support concurrency in three important ways. First, our
analysis is sound for concurrent program executions because
permissions guarantee that program executions are data race free and
reduce thread interactions to specific points in the program such as
forking or joining a thread, or acquiring or releasing a lock.
Second, we develop our analysis for a programming language with
primitives that represent the ownership transfer that happens at these
thread interaction points.  These primitives, $\inhale$ and
$\exhale$ \cite{LeinoMueller'09,MuellerSchwerhoffSummers16}, express
that a thread obtains permissions (for instance, by acquiring a lock)
or loses permissions (for instance, by passing them to another thread
along with a message) and can thereby represent a wide range of thread
interactions in a uniform way~\cite{LeinoMuellerSmans10,SummersMueller18}.  Third,
our analysis distinguishes read and write access and, thus, ensures
exclusive writes while permitting concurrent read accesses. As is
standard, we employ \emph{fractional permissions} \cite{boyland03} for
this purpose; a full permission is required to write to a location,
but any positive fraction permits read access.

\subsubsection{Approach.}

Our analysis reduces the problem of reasoning about permissions for
array elements to reasoning about numerical values for permission
fractions. To achieve this, we represent permission fractions for all
array elements \code{$\a$[$\qi$]} using a \emph{single} numerical expression $t(\a,\qi)$
parameterised by $\a$ and $\qi$. For instance, the conditional term $\cond{\a{=}\code{a}\wedge\qi{=}\code{j}}{1}{0}$
represents full permission (denoted by 1) for array
element \code{a[j]} and no permission for all other array elements.

Our analysis employs a \emph{precise} backwards analysis
for \emph{loop-free} code: a variation on the standard notion
of weakest preconditions. We apply this analysis to
loop bodies to obtain a
permission precondition for a single loop iteration. Per array
element, the \emph{whole loop} requires the \emph{maximum} fraction
over all loop iterations, adjusted by permissions gained
and lost during loop execution.  Rather than computing
permissions via a fixpoint iteration (for which a precise widening
operator is difficult to design), we express them as a maximum over
the variables changed by the loop execution. We then use inferred
numerical invariants on these variables and a novel \emph{maximum
elimination} algorithm to infer a specification for the entire loop.
Permission postconditions are obtained analogously.

\begin{figure}[t]
\centering
\begin{minipage}[c]{0.51\linewidth}
\begin{silver}[mathescape=true]
method copyEven(a: Int[]) {
   var j, v: Int := 0;
   while(j < length(a)) {
      if (j 
      else { a[j] := v };
      j := j + 1
   }
}
\end{silver}
\vspace{-3mm}
\caption{Program \code{copyEven}.}
\label{fig:copyEven}
\end{minipage}
\quad
\begin{minipage}[c]{0.44\linewidth}
\begin{silver}[mathescape=true]
method parCopyEven(a: Int[]) {
   var j: Int := 0;
   while(j < length(a)/2) {
      exhale(a, 2*j, 1/2);
      exhale(a, 2*j+1, 1);
      j := j + 1
   }
}
\end{silver}
\vspace{-3mm}
\caption{Program \code{parCopyEven}.}
\label{fig:parCopyEven}
\end{minipage}
\end{figure}

For the method \code{copyEven} in \figref{copyEven}, the analysis determines that
the permission amount required by a single loop iteration is
$\cond{\dividesTwo{2}{\code{j}}}{\cond{\a{=}\code{a}\mathrel{\wedge}\qi{=}\code{j}}{\rd}{0}}{\cond{\a{=}\code{a}\mathrel{\wedge}\qi{=}\code{j}}{1}{0}}$.
The symbol $\rd$ represents a fractional read permission.
Using a suitable integer invariant for the loop counter \code{j}, we obtain the loop precondition
$\inlinemaxunder{\mathtt{j}\mid 0\leq \mathtt{j}
< \length{\mathtt{a}}}{
\cond{
\dividesTwo{2}{\code{j}}
}{
\cond{\a{=}\code{a}\mathrel{\wedge}\qi{=}\code{j}}{\rd}{0}
}{
\cond{\a{=}\code{a}\mathrel{\wedge}\qi{=}\code{j}}{1}{0}
}
}$.
Our maximum elimination algorithm obtains $\cond{\a{=}\code{a}\mathrel{\wedge} 0{\leq}\qi{<}\length{\code{a}}}{\cond{\dividesTwo{2}{\qi}}{\rd}{1}}{0}$.
By ranging over all $\a$ and $\qi$, this can be read as read permission for even indices and write permission for odd indices within the array \code{a}'s bounds.


\subsubsection{Contributions.}

The contributions of our paper are:
\begin{enumerate}

\item A novel permission inference that uses maximum expressions over
parameterised arithmetic expressions to summarise loops (\secref{wpp} and \secref{loops})

\item An algorithm for eliminating maximum (and minimum) expressions
over an unbounded number of cases (\secref{maxelim})

\item An implementation of our analysis, which will be made available as an artifact

\item An evaluation on benchmark examples from existing papers and
competitions, demonstrating that we obtain sound, precise, and compact
specifications, even for challenging array access patterns and
parallel loops (\secref{evaluation})

\item Proof sketches for the soundness of our permission inference and correctness of our maximum elimination algorithm (included in the appendix.)
\end{enumerate}

%
%
%
%
%

\section{Programming Language}
\label{sec:preliminaries}

\begin{figure}[t]
\[
\begin{array}{rcl}
e &\defequals& n \mid x \mid n{\cdot}x \mid e_1 + e_2 \mid e_1 - e_2 \mid \deref{a}{e} \mid \length{a} \mid \cond{b}{e_1}{e_2}\\
b &\defequals& e_1 \op\; e_2 \mid \divides{n}{e} \mid \notdivides{n}{e} \mid b_1 \wedge b_2 \mid b_1 \vee b_2 \mid \neg b\\
\op &\in& \{=, \neq, <, \leq, >, \geq\}\\
p &\defequals& q \mid \rd \mid p_1 + p_2 \mid p_1 - p_2 \mid \min(p_1,p_2) \mid \max(p_1,p_2) \mid \cond{b}{p_1}{p_2}\\
s &\defequals& \noop \mid \assign{x}{e} \mid \assign{a_1}{a_2} \mid \assign{x}{\deref{a}{e}} \mid \assign{\deref{a}{e_1}}{e_2} \mid \exhale{a,e,p} \mid \inhale{a,e,p}\\
 &\mid& \seq{s_1}{s_2} \mid \ifcond{b}{s_1}{s_2} \mid \while{b}{s}
\end{array}
\]
\vspace{-3mm}
\caption{Programming Language. $n$ ranges over integer
  constants, $x$ over integer variables, $a$ over array
  variables, $q$ over non-negative fractional (permission-typed)
  constants. $e$ stands for integer expressions, and $b$ for
  boolean. Permission expressions $p$ are a
  separate syntactic category.}
\label{fig:language}
\end{figure}

We define our inference technique over the programming language in
\figref{language}. Programs operate on integers (expressions $e$),
booleans (expressions $b$), and one-dimensional integer arrays
(variables $a$); a generalisation to other forms of arrays is straightforward and supported by our implementation.
Arrays are read and updated via the statements
$\assign{x}{\deref{a}{e}}$ and $\assign{\deref{a}{e}}{x}$; array
lookups in expressions are not part of the surface syntax, but are used
internally by our analysis. Permission expressions $p$
evaluate to rational numbers; $\rd$, $\min$,
and $\max$ are for internal use.

A full-fledged programming language contains many statements that
affect the ownership of memory locations, expressed via permissions
\cite{LeinoMuellerSmans10,SummersMueller18}. For example in a concurrent setting, a fork operation may
transfer permissions to the new thread, acquiring a lock obtains
permission to access certain memory locations, and messages may
transfer permissions between sender and receiver. Even in a sequential
setting, the concept is useful: in procedure-modular reasoning,
a method call transfers permissions from the caller to the callee,
and back when the callee terminates. Allocation can be represented as
obtaining a fresh object and then obtaining permission to
its locations.

For the purpose of our permission inference, we can reduce all of
these operations to two basic statements that directly manipulate the
permissions currently held
\cite{LeinoMueller'09,MuellerSchwerhoffSummers16}.  An
$\inhale{a,e,p}$ statement adds the amount $p$ of permission for the
array location $\deref{a}{e}$ to the currently held
permissions. Dually, an $\exhale{a,e,p}$ statement requires that this
amount of permission is \emph{already} held, and then removes it.  We
assume that for any {\inhale} or {\exhale} statements, the permission
expression $p$ denotes a non-negative fraction. For simplicity, we
restrict {\inhale} and {\exhale} statements to a \emph{single} array
location, but the extension to unboundedly-many locations from the
same array is straightforward~\cite{MuellerSchwerhoffSummers16b}.

\subsubsection{Semantics.}

The operational semantics of our language is mostly standard, but is
instrumented with additional state to track how much permission is
held to each heap location; a program state therefore consists of a
triple of heap $H$ (mapping pairs of array identifier and integer
index to integer values), a \emph{permission map} $P$, mapping such
pairs to \emph{permission amounts}, and an environment $`s$ mapping
variables to values (integers or array identifiers).

The execution of $\inhale$ or $\exhale$ statements causes modifications to
the permission map, and all array accesses are guarded with checks
that \emph{at least some} permission is held when reading and that
full ($1$) permission is held when writing~\cite{boyland03}.  If these
checks (or an $\exhale$ statement) fail, the execution terminates with a
\emph{permission failure}. Permission amounts greater than 1
indicate invalid states that cannot be reached by a program execution.
We model run-time errors other than permission failures (in particular, out-of-bounds accesses) as stuck configurations.


\section{Permission Inference for Loop-Free Code}
\label{sec:wpp}

Our analysis infers a sufficient permission precondition and a
guaranteed permission postcondition for each method of a program.
Both conditions are mappings from array elements to permission
amounts.  Executing a statement $s$ in a state whose permission map
$P$ contains at least the permissions required by a
\emph{sufficient permission precondition}
for $s$ is guaranteed to not result in a permission
failure.  A \emph{guaranteed permission postcondition} expresses the
permissions that will at least be held when $s$ terminates (see~\appsecref{auxinference}
for formal definitions).

In this section, we define inference rules to compute sufficient
permission preconditions for loop-free code. For programs which do not
add or remove permissions via $\inhale$ and $\exhale$ statements, the
same permissions will still be held after executing the code; however,
to infer guaranteed permission postconditions in the general case, we
also infer the difference in permissions between the state
before and after the execution. We will discuss loops in the next
section. Non-recursive method calls can be handled by applying our
analysis bottom-up in the call graph and using $\inhale$ and $\exhale$
statements to model the permission effect of calls. Recursion can be
handled similarly to loops, but is omitted here.

\begin{figure}[t]
\[
\begin{array}{c}
%
%
\begin{array}{rclrcl}
\wpp{\noop}{p} & = & p \quad &
\wpp{\seq{s_1}{s_2}}{p} & = & \wpp{s_1}{\wpp{s_2}{p}} \\
\wpp{\assign{x}{e}}{p} & = & \sub{p}{e}{x} \quad &
\wpp{\assign{x}{\deref{a}{e}}}{p} & = & \max(\sub{p}{\deref{a}{e}}{x}, \accperm{a}{e}{\rd})\\
\wpp{\assign{\deref{a}{e}}{x}}{p} & = & \multicolumn{4}{l}{\max(\psub{p}{a'}{e'}{\cond{e=e'\wedge a=a'}{x}{\deref{a'}{e'}}}, \accperm{a}{e}{1})}
\end{array} \\
\begin{array}{rclrcl}
\wpp{\exhale{a,e,p'}}{p} & = & p + \accperm{a}{e}{p'} \quad &
\wpp{\inhale{a,e,p'}}{p} & = & \max(0, p - \accperm{a}{e}{p'})
\end{array} \\
\wpp{\ifcond{b}{s_1}{s_2}}{p} = \cond{b}{\wpp{s_1}{p}}{\wpp{s_2}{p}} \\[0.5em]
%
%
\begin{array}{rclrcl}
\diff{\noop}{p} & = & p \quad &
\diff{\seq{s_1}{s_2}}{p} & = & \diff{s_1}{\diff{s_2}{p}} \\
\diff{\assign{x}{e}}{p} & = & \sub{p}{e}{x} \quad &
\diff{\assign{x}{\deref{a}{e}}}{p} & = & \sub{p}{\deref{a}{e}}{x} \\
\diff{\assign{\deref{a}{e}}{x}}{p} & = & \multicolumn{4}{l}{\psub{p}{a'}{e'}{\cond{e=e'\wedge a=a'}{x}{\deref{a'}{e'}}}} \\
\end{array} \\
\begin{array}{rclrcl}
\diff{\exhale{a,e,p'}}{p} & = & p - \accperm{a}{e}{p'} \quad &
\diff{\inhale{a,e,p'}}{p} & = & p + \accperm{a}{e}{p'}
\end{array} \\
\diff{\ifcond{b}{s_1}{s_2}}{p} = \cond{b}{\diff{s_1}{p}}{\diff{s_2}{p}}
\end{array}
\]
\caption{The backwards analysis rules for permission preconditions and relative permission differences.
The notation $\accperm{a}{e}{p}$ is a shorthand for
$\cond{\a{=}a\mathrel{\wedge} \qi{=}e}{p}{0}$ and denotes $p$
permission for the array location $\deref{a}{e}$. Moreover,
$\psub{p}{a'}{e'}{e}$ matches all array accesses in $p$ and replaces
them with the expression obtained from $e$ by substituting all
occurrences of $a'$ and $e'$ with the matched array and index,
respectively. The cases for inhale statements are slightly simplified;
the full rules are given in \appfigref{full-wpp-delta}.}
\label{fig:wpp-delta}
\end{figure}

We define our permission analysis to track and generate
\emph{permission expressions} parameterised by two distinguished
  variables $\a$ and $\qi$; by parameterising our expressions in this
way, we can use a single expression to represent a permission amount
for each pair of $\a$ and $\qi$ values.

\subsubsection{Preconditions.}

The \emph{permission precondition} of a loop-free statement $s$
and a postcondition permission $p$ (in which $\a$ and $\qi$
potentially occur) is denoted by $\wpp{s}{p}$, and is defined in
\figref{wpp-delta}. Most rules are straightforward adaptations
of a classical weakest-precondition computation. Array lookups require
some permission to the accessed array location; we use the internal
expression $\rd$ to denote a non-zero permission amount; a
post-processing step can later replace $\rd$ by a concrete
rational. Since downstream code may require further permission for
this location, represented by the permission expression
$p$, we take the maximum of both amounts. Array updates require full
permission and need to take aliasing into account. The case for
$\inhale$ subtracts the inhaled permission amount from the permissions
required by downstream code; the case for $\exhale$ adds
the permissions to be exhaled. Note that this addition may lead to a
required permission amount exceeding the full permission. This
indicates that the statement is not feasible, that is, all executions
will lead to a permission failure.


To illustrate our $\wpp$ definition, let $s$ be
the body of the loop in the \code{parCopyEven}
method in \figref{parCopyEven}. The precondition $\wpp{s}{0} =
\cond{\a{=}\code{a}\mathrel{\wedge}\qi=2{*}\code{j}}{1/2}{0}+\cond{\a{=}\code{a}\mathrel{\wedge}\qi{=}2{*}\code{j}{+}1}{1}{0}$ expresses that a loop iteration
requires a half permission for the even elements of array \code{a} and
full permission for the odd elements.

\subsubsection{Postconditions.}

The final state of a method execution includes the permissions held in
the method pre-state, adjusted by the permissions that are inhaled or
exhaled during the method execution. To perform this adjustment, we
compute the difference in permissions before and after executing a
statement. The \emph{relative permission difference} for a loop-free
statement $s$ and a permission expression $p$ (in which $\a$ and $\qi$
potentially occur) is denoted by $\diff{s}{p}$, and is defined backward,
analogously to $\wpp$ in \figref{wpp-delta}. The second parameter
$p$ acts as an accumulator; the difference in permission is
represented by evaluating $\diff{s}{0}$.


For a statement $s$ with precondition $\wpp{s,0}$, we obtain the
postcondition $\wpp{s,0} + \diff{s,0}$.
Let $s$ again be the loop body from \code{parCopyEven}.  Since $s$
contains \code{exhale} statements, we obtain $\diff{s,0} =
0-\cond{\a{=}\code{a}\mathrel{\wedge}\qi{=}2{*}\code{j}}{1/2}{0}-\cond{\a{=}\code{a}\mathrel{\wedge}\qi{=}2{*}\code{j}{+}1}{1}{0}$.
Thus, the postcondition $\wpp{s,0} + \diff{s,0}$ can be simplified to $0$. This
reflects the fact that all required permissions for a single loop
iteration are lost by the end of its execution.

Since our $\diff$ operator performs a backward analysis, our permission
post-conditions are expressed in terms of the pre-state of the
execution of $s$. To obtain classical postconditions, any heap
accesses need to refer to the pre-state heap, which can be achieved in
program logics by using \code{old} expressions or logical variables.
Formalizing the postcondition inference as a backward analysis simplifies
our treatment of loops and has technical advantages over classical
strongest-postconditions, which introduce existential
quantifiers for assignment statements. A limitation of our approach
is that our postconditions cannot capture situations in which a statement
obtains permissions to locations for which no pre-state expression
exists, \eg{} allocation of new arrays. Our postconditions
are sound; to make them precise for such cases, our
inference needs to be combined with an additional forward analysis,
which we leave as future work.

\section{Handling Loops via Maximum Expressions}
\label{sec:loops}

In this section, we first focus on obtaining a sufficient permission
precondition for the execution of a loop in isolation (independently
of the code after it) and then combine the inference for loops
with the one for loop-free code described above.

\subsection{Sufficient Permission Preconditions for Loops}
A sufficient permission precondition for a loop guarantees the absence of permission failures for a potentially unbounded number of executions of the loop body. This concept is different from a loop invariant: we require a precondition for all executions of a particular loop, but it need not be inductive. Our technique obtains such a loop precondition by projecting a permission precondition for a single loop iteration over all possible initial states for the loop executions.

\subsubsection{Exhale-Free Loop Bodies.}

We consider first the simpler (but common) case of a loop that does not
contain $\exhale$ statements, \eg{}, does not transfer permissions
to a forked thread. The solution for this case is also
sound for loop bodies where each $\exhale$ is followed by an $\inhale$
for the same array location and at least the same permission amount,
as in the encoding of most method calls.

Consider a sufficient permission precondition $p$ for the body of a
loop $\while{b}{s}$. By definition, $p$ will denote sufficient
permissions to execute $s$ once; the precise locations to which $p$
requires permission depend on the initial state of the loop
iteration. For example, the sufficient permission precondition
for the body of the \code{copyEven} method in \figref{copyEven},
$\cond{
\dividesTwo{2}{\code{j}}
}{\cond{\a{=}\code{a}\mathrel{\wedge}\qi{=}\code{j}}{\rd}{0}}{\cond{\a{=}\code{a}\mathrel{\wedge}\qi{=}\code{j}}{1}{0}}$,
requires permissions to different array locations, depending on the
value of \code{j}. To obtain a sufficient permission precondition for
the entire loop, we leverage an \emph{over-approximating} loop invariant
$\Invplus$ from an off-the-shelf numerical analysis (e.g.,
\cite{CousotH78}) to over-approximate all possible values of the numerical
variables that get assigned in the loop body, here, \code{j}.
We can
then express the loop precondition using the \emph{pointwise maximum}
$\inlinemaxunder{\mathtt{j}\mid \Invplus\wedge b}{p}$, over the values of
\code{j} that satisfy the condition $\Invplus\wedge b$. (The maximum
over an empty range is defined to be $0$.)
For the
\code{copyEven} method, given the invariant $0\leq \code{j} \leq \length{\code{a}}$, the loop precondition is
$\inlinemaxunder{\mathtt{j}\mid 0 \leq \mathtt{j} <
  \length{\mathtt{a}}}{p}$.

In general, a permission precondition for a loop body may also depend
on array \emph{values}, \eg{}, if those values are used in branch
conditions. To avoid the need for an expensive array value analysis,
we define both an over- and an under-approximation of permission expressions, denoted $\overapprox{p}$ and $\underapprox{p}$ (cf.~%
\appsecref{restrictionapp}), with the guarantees that $p \leq \overapprox{p}$ and $\underapprox{p} \leq p$. These approximations abstract away array-dependent conditions, and have
%
 an impact on precision only when array values
are used to determine a location to be accessed. For example, a
linear array search for a particular value accesses the array only up to the
(a-priori unknown) point at which the value is found, but our
permission precondition conservatively requires access to the full array.

\begin{theorem}\label{thm:stable}
Let $\while{b}{s}$ be an exhale-free loop, let
$\Vec{x}$ be the integer variables modified by $s$, and let
$\Invplus$ be a sound over-approximating numerical loop invariant
(over the integer variables in $s$). Then $\inlinemaxunder{\Vec{x}
  \mid \Invplus \wedge b}{\restrict{\wpp{s}{0}}}$ is a sufficient
permission precondition for $\while{b}{s}$.
\end{theorem}

\subsubsection{Loops with Exhale Statements.}

For loops that contain $\exhale$ statements, the
approach described above does not always guarantee a sufficient
permission precondition. For example, if a loop gives away full
permission to the \emph{same} array location in every iteration,
our pointwise maximum construction yields a precondition requiring the
full permission once, as opposed to the \emph{unsatisfiable}
precondition (since the loop is guaranteed to cause a
permission failure).

As explained above, our inference is sound if each $\exhale$ statement
is followed by a corresponding $\inhale$, which can often be checked
syntactically. In the following, we present another decidable condition that
guarantees soundness and that can be checked efficiently by an SMT solver. If
neither condition holds, we preserve soundness by inferring an unsatisfiable
precondition; we did not encounter
any such examples in our evaluation.

Our soundness condition checks that the maximum of the permissions
required by two loop iterations is not less than the permissions
required by executing the two iterations in sequence. Intuitively,
that is the case when neither iteration removes permissions that are
required by the other iteration.


\begin{theorem}[Soundness Condition for Loop Preconditions]\label{thm:condition}
Given a loop $\while{b}{s}$, let $\Vec{x}$ be the integer variables
modified in $s$ and let $\Vec{v}$ and $\Vec{v'}$ be two fresh sets of
variables, one for each of $\Vec{x}$. Then $\inlinemaxunder{\Vec{x}
  \mid \Invplus \wedge b}{\restrict{\wpp{s}{0}}}$ is a sufficient
permission precondition for $\while{b}{s}$ if the following implication
is valid in all states:
\[
\begin{array}{c}
\vecsub{(\Inv\wedge b)}{{v}}{{x}}\ \wedge\ \vecsub{(\Inv\wedge b)}{{v'}}{{x}}\ \wedge\ (\bigvee{\Vec{v\neq v'}})\ \Rightarrow\\
 \max(\vecsub{\restrict{\wpp{s,0}}}{{v}}{{x}}, \vecsub{\restrict{\wpp{s,0}}}{{v'}}{{x}}) \geq \vecsub{\restrict{\wpp{s}{\vecsub{\restrict{\wpp{s}{0}}}{{v'}}{{x}}}}}{{v}}{{x}}\end{array}
\]
\end{theorem}
The additional variables $\Vec{v}$ and $\Vec{v'}$ are used to model
two arbitrary valuations of $\Vec{x}$; we constrain these to represent
two initial states allowed by $\Inv\wedge b$ and
different from each other for at least one program variable. We
then require that the effect of analysing each loop iteration
independently and taking the maximum is not smaller than the effect of
sequentially composing the two loop iterations.

The theorem requires implicitly that no two different iterations of a
loop observe exactly the same values for all integer
variables. If that could be the case, the condition $\bigvee{\Vec{v\neq v'}}$ would cause us to ignore a potential pair of initial states for two different loop iterations. To avoid
this problem, we assume that all loops satisfy this requirement; it
can easily be enforced by adding an additional variable as loop
iteration counter \cite{GuptaR09}.

For the \code{parCopyEven} method (\figref{parCopyEven}), the
soundness condition holds since, due to the $v\neq v'$ condition, the
two terms on the right of the implication are equal for all values of $\qi$.
We can thus infer a sufficient precondition as
$\inlinemaxunder{\mathtt{j}\mid 0\leq \mathtt{j}
< \length{\mathtt{a}}/2}{
\cond{\a{=}\code{a}\mathrel{\wedge}\qi=2{*}\code{j}}{1/2}{0}+\cond{\a{=}\code{a}\mathrel{\wedge}\qi{=}2{*}\code{j}{+}1}{1}{0}
}$.



\subsection{Permission Inference for Loops}

We can now extend the pre- and postcondition inference from
\secref{wpp} with loops. $\wpp{\while{b}{s}}{p}$ must require
permissions such that (1)~the loop executes without permission failure
and (2)~at least the permissions described by $p$ are held when the
loop terminates. While the former is provided by the loop precondition
as defined in the previous subsection, the latter also depends on the
permissions gained or lost during the execution of the loop. To characterise
these permissions, we extend the $\diff$ operator from \secref{wpp}
to handle loops.

Under the soundness condition from \thmref{condition}, we can mimic
the approach from the previous subsection and use over-approximating
invariants to project out the permissions \emph{lost} in a single loop
iteration (where $\diff{s}{0}$ is negative) to those lost by the entire loop, using a maximum
expression. This projection conservatively assumes that the
  permissions lost in a single iteration are lost by all
  iterations whose initial state is allowed by the loop invariant and
loop condition. This approach is a sound over-approximation of the
permissions \emph{lost}.

However, for the permissions \emph{gained} by a loop
iteration (where $\diff{s}{0}$ is positive), this approach would be unsound because the
over-approximation includes iterations that may not actually happen
and, thus, permissions that are not actually gained.
For this reason, our technique handles gained permissions
 via an
\emph{under-approximate}\footnote{An under-approximate loop invariant
  must be true \emph{only} for states that will actually be
  encountered when executing the loop.} numerical loop
invariant $\Invminus$ (e.g., \cite{Mine12}) and thus projects
the gained permissions only over iterations that will surely happen.

This approach is reflected in the definition of our $\diff$
  operator below via $d$, which represents the permissions \emph{possibly
    lost} or \emph{definitely gained} over all iterations of the loop. In the former case,
we have $\diff{s}{0} < 0$ and, thus, the first summand is 0
and the computation based on the over-approximate invariant applies
(note that the negated maximum of negated values is the minimum; we take the
minimum over negative values).
In the latter case ($\diff{s}{0} > 0$), the second summand is 0 and
the computation based on the under-approximate invariant applies
(we take the maximum over positive values).
\[
\begin{array}{rcl}
\diff{\while{b}{s}}{p} &=& \cond{b}{d + p'}{p}\textit{, where:}\\
d &=& \maxunder{\Vec{x} \mid \Invminus \wedge b}{\underapprox{\max(0,\diff{s}{0})}} - \maxunder{\Vec{x} \mid \Invplus \wedge b}{\overapprox{\max(0,-\diff{s}{0})}}\\
p' &=& \maxunder{\Vec{x} \mid \Invminus \wedge \neg b}{\underapprox{\max(0,p)}} - \maxunder{\Vec{x} \mid \Invplus \wedge \neg b}{\overapprox{\max(0,-p)}}\\\end{array}
\]
$\Vec{x}$ denotes again the integer variables modified in $s$.
The role of $p'$ is to carry over the permissions $p$ that are gained
or lost by the code following the loop, taking into account any state
changes performed by the loop.
Intuitively, the maximum expressions replace the variables
  $\Vec{x}$ in $p$ with expressions that do not
  depend on these variables but nonetheless reflect properties of their values right
  after the execution of the loop. For permissions gained, these properties
are based on the under-approximate loop
invariant to ensure that they hold for any possible loop execution. For permissions lost,
we use the over-approximate invariant.
For the loop in  \code{parCopyEven} we use the  invariant
$0\leq \mathtt{j} \leq \length{\mathtt{a}}/2$ to obtain
\mbox{$d=-\inlinemaxunder{\mathtt{j}\mid 0\leq \mathtt{j}
< \length{\mathtt{a}}/2}{
\cond{\a{=}\code{a}\mathrel{\wedge}\qi=2{*}\code{j}}{1/2}{0}+\cond{\a{=}\code{a}\mathrel{\wedge}\qi{=}2{*}\code{j}{+}1}{1}{0}
}$}.
Since there are no statements following the loop, $p$ and therefore $p'$ are 0.

Using the same $d$ term, we can now define the general case of $\wpp$
for loops, combining (1)~the loop precondition and (2)~the permissions
required by the code after the loop, adjusted by the permissions
gained or lost during loop execution:
\[
\wpp{\while{b}{s}}{p} = \cond{b}{\max(\maxunder{\Vec{x} \mid \Invplus \wedge b}{\restrict{\wpp{s}{0}}},\maxunder{\Vec{x} \mid \Invplus \wedge \neg b}{(\restrict{p})}-d)}{p}
\]
Similarly to $p'$ in the rule for $\diff$, the expression $\inlinemaxunder{\Vec{x} \mid \Invplus \wedge \neg b}{\restrict{p}}$ conservatively over-approximates the permissions required to execute the code after the loop.
For method \code{parCopyEven}, we obtain a sufficient precondition
that is the negation of the $\diff$. Consequently, the postcondition is 0.

\subsubsection{Soundness.}
Our $\wpp$ and $\diff$ definitions
yield a sound method for computing sufficient permission preconditions
and guaranteed postconditions:
\begin{theorem}[Soundness of Permission Inference]\label{thm:soundness}
For any statement $s$, if every $\whilesym$ loop in $s$ either is
exhale-free or satisfies the condition of \thmref{condition} then
$\wpp{s}{0}$ is a sufficient permission precondition for $s$, and
$\wpp{s}{0} + \diff{s}{0}$ is a corresponding guaranteed permission
postcondition.
\end{theorem}

Our inference expresses pre and postconditions using a
maximum operator over an unbounded set of values.
However, this operator is not supported by
SMT solvers. To be able to use the inferred conditions
for SMT-based verification, we provide an algorithm for eliminating these
operators, as we discuss next.

\section{A Maximum Elimination Algorithm}
\label{sec:maxelim}

We now present a new algorithm for replacing maximum expressions
over an unbounded set of values (called \emph{pointwise maximum expressions}
in the following) with equivalent expressions containing no pointwise maximum expressions.
Note that, technically our algorithm computes solutions to $\max_{x \suchthat b \wedge p \geq 0}(p)$ since some optimisations exploit the fact that the permission expressions our analysis generates always denote non-negative values.

\subsection{Background: Quantifier Elimination}

Our algorithm builds upon ideas from Cooper's classic \emph{quantifier elimination} algorithm \cite{CooperQE} which, given a formula $\exists x. b$ (where $b$ is a quantifier-free Presburger formula), computes an equivalent quantifier-free formula $b'$. Below, we give a brief summary of Cooper's approach.

The problem is first reduced via boolean and arithmetic manipulations to a formula $\exists x.b$ in which $x$ occurs at most once per literal and with no coefficient.
The key idea is then to reduce $\exists x.b$ to a disjunction of two cases: (1) there is a \emph{smallest} value of $x$ making $b$ true, or (2) $b$ is true for \emph{arbitrarily small} values of $x$.

In case (1), one computes a \emph{finite} set of expressions $S$ (the $b_i$ in \cite{CooperQE}) guaranteed to include the smallest value of $x$. For each (in/dis-)equality literal containing $x$ in $b$, one collects a \emph{boundary expression} $e$ which denotes a value for $x$ making the literal true, while the value $e-1$ would make it false. For example, for the literal $y < x$ one generates the expression $y+1$. If there are no (non-)divisibility constraints in $b$, by definition, $S$ will include the smallest value of $x$ making $b$ true.
To account for (non-)divisibility constraints such as $\divides{2}{x}$, the lowest-common-multiple $`d$ of the divisors (and $1$) is returned along with $S$; the guarantee is then that the smallest value of $x$ making $b$ true will be $e+d$ for some $e\in S$ and $d\in\range{0}{`d-1}$. We use $\mini{b}$ to denote the function handling this computation. Then, $\exists x.b$ can be reduced to $\bigvee_{e \in S, d \in \range{0}{`d -1}}\sub{b}{e + d}{x}$, where $(S,`d) = \mini{b}$.

In case (2), one can observe that the (in/dis-)equality literals in $b$ will flip value at finitely many values of $x$, and so for \emph{sufficiently small} values of $x$, \emph{each} (in/dis-)equality literal in $b$ will have a constant value (\eg{}, $y > x$ will be $\trueExp$). By replacing these literals with these constant values, one obtains a new expression $b'$ equal to $b$ for small enough $x$, and which depends on $x$ only via (non-)divisibility constraints. The value of $b'$ will therefore actually be determined by $x\mod `d$, where $`d$ is the lowest-common-multiple of the (non-)divisibility constraints. We use $\leftInf{b}$ to denote the function handling this computation. Then, $\exists x.b$ can be reduced to $\bigvee_{d \in \range{0}{`d -1}}\sub{b'}{d}{x}$, where $(b',`d) = \leftInf{b}$.

In principle, the maximum of a function $y = \max_x f(x)$ can be defined using two first-order quantifiers $\forall x. f(x) \leq y$ and $\exists x. f(x) = y$. One might therefore be tempted to tackle our maximum elimination problem using quantifier elimination directly. We explored this possibility and found two serious drawbacks. First, the resulting formula does not yield a permission-typed expression that we can plug back into our analysis. Second, the resulting formulas are extremely large (\eg{}, for the \code{copyEven} example it yields several pages of specifications), and hard to simplify since relevant information is often spread across many terms due to the two separate quantifiers. Our maximum elimination algorithm addresses these drawbacks by natively working with arithmetic expression, while mimicking the basic ideas of Cooper's algorithm and incorporating domain-specific optimisations.

\subsection{Maximum Elimination}\label{sec:simple}


The first step is to reduce the problem of eliminating general $\inlinemaxunder{x \suchthat b}{p}$ terms to those in which $b$ and $p$ come from a simpler restricted grammar. These \emph{simple permission expressions} $p$ do not contain general conditional expressions $\cond{b'}{p_1}{p_2}$, but instead only those of the form $\leaf{b'}{r}$ (where $r$ is a constant or $\rd$). Furthermore, simple permission expressions only contain subtractions of the form $p - \cond{b'}{r}{0}$. This is achieved in a precursory rewriting of the input expression by, for instance, distributing pointwise maxima over conditional expressions and binary maxima. For example, 
the pointwise maximum term (part of the \code{copyEven} example): $\inlinemaxunder{\mathtt{j} \suchthat 0\leq \mathtt{j} < \length{\mathtt{a}}}{\cond{\divides{2}{\code{j}}}{\cond{\a{=}\code{a}\wedge\qi{=}\code{j}}{\rd}{0}}{\cond{\a{=}\code{a}\wedge\qi{=}\code{j}}{1}{0}}}$\\will be reduced to:
\[\begin{array}{ll}\max(&\inlinemaxunder{\mathtt{j} \suchthat 0\leq \mathtt{j} < \length{\mathtt{a}} \wedge \divides{2}{\mathtt{j}}}{\cond{\a{=}\code{a}\wedge\qi{=}\code{j}}{\rd}{0}},\\
&\inlinemaxunder{\mathtt{j} \suchthat 0\leq \mathtt{j} < \length{\mathtt{a}} \wedge \notdivides{2}{\mathtt{j}}}{\cond{\a{=}\code{a}\wedge\qi{=}\code{j}}{1}{0}})\end{array}\]

\subsubsection{Arbitrarily-small Values.}

We exploit a high-level case-split in our algorithm design analogous to Cooper's: given a pointwise maximum expression $\inlinemaxunder{x \suchthat b}{p}$, either a \emph{smallest} value of $x$ exists such that $p$ has its maximal value (and $b$ is true), or there are \emph{arbitrarily small} values of $x$ defining this maximal value. To handle the arbitrarily small case, we define a completely analogous $\leftInf{p}$ function, which recursively replaces all boolean expressions $b'$ in $p$ with $\leftInf{b'}$ as computed by Cooper; we relegate the definition to \appsecref{leftinf}.
We then use $\cond{b'}{p'}{0}$, where $(b', \delta_1) = \leftInf{b}$ and $(p', \delta_2) = \leftInf{p}$, as our expression in this case. Note that this expression still depends on $x$ if it contains (non-)divisibility constraints; \thmref{maxelim} shows how $x$ can be eliminated using $\delta_1$ and~$\delta_2$.
%

\subsubsection{Selecting Boundary Expressions for Maximum Elimination.}\label{sec:minimax}
Next, we consider the case of selecting an appropriate set of boundary expressions, given a $\maxunder{x \suchthat b}{(p)}$ term. We define this first for $p$ in isolation, and then give an extended definition accounting for the $b$. Just as for Cooper's algorithm, the boundary expressions must be a set guaranteed to include the \emph{smallest} value of $x$ defining the maximum value in question. The set must be finite, and be as small as possible for efficiency of our overall algorithm. We refine the notion of boundary expression, and compute a set of \emph{pairs} $(e,b')$ of integer expression $e$ and its \emph{filter condition} $b'$: the filter condition represents an additional condition under which $e$ must be included as a boundary expression. In particular, in contexts where $b'$ is false, $e$ can be ignored; this gives us a way to symbolically define an ultimately-smaller set of boundary expressions, particularly in the absence of contextual information which might later show $b'$ to be false. We call these pairs \emph{filtered boundary expressions}.
\begin{figure}[t]
\[
\begin{array}{l}
\begin{array}{rcl}
\minimax{\leaf{b}{p}} &=& (T,\delta),\textit{ where }(S,\delta)=\mini{b}, T=\{(e,\trueexp)\mid e \in S\} \\
\minimax{p_1 + p_2} &=& (T_1\union T_2, \lcm{`d_1,`d_2})\\
\textit{where}\quad(T_1,`d_1) &=& \minimax{p_1}, \ (T_2,`d_2) = \minimax{p_2}\\
\minimax{\max(p_1,p_2)} &=& \minimax{\min(p_1,p_2)} = \minimax{p_1 + p_2} \textit{ as above}\\
\minimax{p_1 - \leaf{b}{p}} &=& (T_1\union T_2, \lcm{`d_1,`d_2})\\
\textit{where}\quad(T_1,`d_1) &=& \minimax{p_1}, \ (S_2,`d_2) = \mini{\neg b},\\
T_2' &=& \{(e, p_1>0) \mid\ e \in S_2\}\end{array}\\\ \\
\begin{array}{rcl}
\minimax{(p,b)} &=& (T_p\union T_b', `d')\textit{ where }(T_p,`d_p) = \minimax{p},\ (S_b,`d_b) = \mini{b},\\
\multicolumn{3}{l}{\ `d' = \lcm{`d_p,`d_b},\ (b',`d_b) = \leftInf{b},\ (p',`d_p) = \leftInf{p},}\\
\multicolumn{3}{l}{T_b' = \{(e_b,(\kern-3pt\veeunder{d\in\range{0}{`d'-1}}{\kern-10pt(\sub{(\neg b'\wedge p'>0)}{d}{x})})\vee\kern-6pt\veeunder{\substack{(e_p,b_p)\in T_p\\d_p\in\range{0}{`d_p-1}}}{\kern-10pt\sub{(\neg b\wedge b_p)}{(e_p + d_p)}{x}}) \;\mid\; e_b \in S_b\}}\\
\end{array}
\end{array}
\]
\vspace{-3mm}
\caption{Filtered boundary expression computation.}\label{fig:miniPerm}
\end{figure}
\begin{definition}[Filtered Boundary Expressions]\label{def:mini}
The \emph{filtered boundary expression computation for $x$ in $\simple{p}{x}$}, written $\minimax{\simple{p}{x}}$, returns a pair of a set $T$ of pairs $(e,b')$, and an integer constant $`d$, as defined in \figref{miniPerm}. This definition is also overloaded with a definition of \emph{filtered boundary expression computation for $(x \suchthat \simple{b}{x})$ in $\simple{p}{x}$}, written $\minimax{(\simple{p}{x},\simple{b}{x})}$.
\end{definition}
Just as for Cooper's $\mini{b}$ computation, our function $\minimax{p}$ computes the set $T$ of $(e,b')$ pairs along with a single integer constant $`d$, which is the least common multiple of the divisors occurring in $p$; the desired smallest value of $x$ may actually be some $e+d$ where $d\in\range{0}{`d-1}$. There are three key points to \defref{mini} which ultimately make our algorithm efficient:

First, the case for $\minimax{\leaf{b}{p}}$ only includes boundary expressions for making $b$ \emph{true}. The case of $b$ being false (from the structure of the permission expression) is not relevant for trying to maximise the permission expression's value (note that this case will never apply under a subtraction operator, due to our simplified grammar, and the case for subtraction not recursing into the right-hand operand).

Second, the case for $\minimax{p_1 - \leaf{b}{p}}$ dually only considers boundary expressions for making $b$ \emph{false} (along with the boundary expressions for maximising $p_1$). The filter condition $p_1 > 0$ is used to drop the boundary expressions for making $b$ false; in case $p_1$ is not strictly positive we know that the evaluation of the whole permission expression will not yield a strictly-positive value, and hence is not an interesting boundary value for a non-negative maximum.

Third, in the overloaded definition of $\minimax{(p,b)}$, we combine boundary expressions for $p$ with those for $b$. The boundary expressions  for $b$ are, however, superfluous \emph{if}, in analysing $p$ we have already determined a value for $x$ which maximises $p$ and happens to satisfy $b$. If all boundary expressions for $p$ (whose filter conditions are true) make $b$ true, \emph{and} all non-trivial (\ie{} strictly positive) evaluations of $\leftInf{p}$ used for potentially defining $p$'s maximum value also satisfy $b$, then we can safely discard the boundary expressions for $b$.

We are now ready to reduce pointwise maximum expressions to equivalent maximum expressions over finitely-many cases:
\begin{theorem}[Simple Maximum Expression Elimination]\label{thm:maxelim}
For any pair $(\simple{p}{x},\simple{b}{x})$, if $\entails p \geq 0$, then we have:
\begin{equation*}
\begin{split}
\entails \maxunder{x \suchthat b}{p} = \max\bigl(\maxunder{\substack{(e,b'') \in T \\ d \in \range{0}{\delta-1}}}{\cond{b'' \wedge \sub{b}{e + d}{x}}{\sub{p}{e + d}{x}}{\zeroPerm}}), \\[-10pt]
\maxunder{d \in \range{0}{\lcm{`d_1,`d_2}-1}}{\cond{\sub{b'}{d}{x}}{\sub{p'}{d}{x}}{\zeroPerm}} \bigr)
\end{split}
\end{equation*}
where $(T,`d) = \minimax{(p,b)}$, $(b',`d_1) = \leftInf{b}$ and $(p',`d_2) = \leftInf{p}$.
\end{theorem}

To see how our filter conditions help to keep the set $T$ (and therefore, the first iterated maximum on the right of the equality in the above theorem) small, consider the example: $\inlinemaxunder{x \suchthat x{\geq}0}{\leaf{x{=}i}{1}}$ (so $p$ is $\leaf{x{=}i}{1}$, while $b$ is $x \geq 0$). In this case, evaluating $\minimax{(p,b)}$ yields the set $T = \{(i, \trueval), (0, i<0)\}$ with the meaning that the boundary expression $i$ is considered in all cases, while the boundary expression $0$ is only of interest if $i < 0$. The first iterated maximum term would be
\mbox{$\max(\cond{\trueval \wedge i{\geq}0}{\leaf{i{=}i}{1}}{0}, \cond{i{<}0 \wedge 0{\geq}0}{\leaf{0{=}i}{1}}{0})$}.
We observe that the term corresponding to the boundary value $0$ can be simplified to $0$ since it contains the two contradictory conditions $i < 0$ and $0 = i$. Thus, the entire maximum can be simplified to $\cond{i{\geq}0}{1}{0}$. Without the filter conditions the result would instead be $\max(\cond{i{\geq}0}{1}{0}, \cond{0{=}i}{1}{0})$. In the context of our permission analysis, the filter conditions allow us to avoid generating boundary expressions corresponding \eg{} to the integer loop invariants, provided that the expressions generated by analysing the permission expression in question already suffice. We employ aggressive syntactic simplification of the resulting expressions, in order to exploit these filter conditions to produce succinct final answers.

\section{Implementation and Experimental Evaluation}
\label{sec:evaluation}

\newcommand{\yes}{\textcolor{green!75!black}{\ding{51}\xspace}}
\newcommand{\very}{\yes*}
\newcommand{\no}{\textcolor{red}{\ding{55}\xspace}}

\begin{table}[t]
\begin{footnotesize}
	\begin{center}
		\begin{tabular}{p{5.6em}ccccc}
			\toprule
			Program & LOC & Loops & Size & Prec. & Time \\
			\midrule
			\texttt{addLast}           & 12 & 1 (1) & 1.9  & \yes  & 21 \\ 
			\texttt{append}            & 13 & 1 (1) & 1.9  & \yes  & 32 \\ 
			\texttt{array1}            & 17 & 2 (2) & 0.9  & \no   & 28 \\ 
			\texttt{array2}            & 23 & 3 (2) & 0.9  & \no   & 35 \\ 
			\texttt{array3}            & 23 & 2 (2) & 1.1  & \yes  & 24 \\ 
			\texttt{arrayRev}          & 18 & 1 (1) & 3.2  & \very & 28 \\ 
			\texttt{bubbleSort}        & 23 & 2 (2) & 1.8  & \very & 34 \\ 
			\texttt{copy}              & 16 & 2 (1) & 1.6  & \yes  & 27 \\ 
			\texttt{copyEven}         & 17 & 1 (1) & 1.6  & \yes  & 27 \\ 
			\texttt{copyEven2}         & 14 & 1 (1) & 1.4  & \no   & 20 \\ 
			\texttt{copyEven3}         & 14 & 1 (1) & 2.2  & \very & 23 \\ 
			\texttt{copyOdd}           & 21 & 2 (1) & 2.4  & \yes  & 55 \\ 
			\texttt{copyOddBug}        & 19 & 2 (1) & 7.1  & \yes  & 57 \\ 
			\texttt{copyPart}          & 17 & 2 (1) & 1.7  & \yes  & 30 \\ 
			\texttt{countDown}         & 21 & 3 (2) & 1.1 & \yes  & 32 \\ 
			\texttt{diff}              & 31 & 2 (2) & 2.0  & \no   & 70 \\ 
			\texttt{find}              & 19 & 1 (1) & 3.0  & \yes  & 43 \\ 
			\texttt{findNonNull}       & 19 & 1 (1) & 3.0   & \yes  & 40 \\ 
			\texttt{init}              & 18 & 2 (1) & 1.1  & \yes  & 28 \\ 
			\texttt{init2d}            & 23 & 2 (2) & 2.1  & \yes  & 52 \\ 
			\texttt{initEven}          & 18 & 2 (1) & 0.9  & \no   & 26 \\ 
			\texttt{initEvenbug}       & 18 & 2 (1) & 1.5  & \no   & 28 \\ 
			\texttt{initNonCnst}       & 18 & 2 (1) & 1.1  & \yes  & 27 \\ 
			\texttt{initPart}          & 19 & 2 (1) & 1.1  & \yes  & 30 \\ 
			\bottomrule
		\end{tabular}
		\hspace{0em}
		\begin{tabular}{p{5.6em}ccccc}
			\toprule
			Program & LOC & Loops & Size & Prec. & Time \\
			\midrule
			\texttt{initPartBug}       & 19 & 2 (1) & 1.5  & \yes  & 31 \\ 
			\texttt{insertSort}        & 21 & 2 (2) & 2.5  & \very & 35 \\ 
			\texttt{javaBubble}        & 24 & 2 (2) & 2.3  & \very & 32 \\ 
			\texttt{knapsack}          & 21 & 2 (2) & 1.3  & \no   & 45 \\ 
			\texttt{lis}               & 37 & 4 (2) & 4.2  & \yes  & 73 \\ 
			\texttt{matrixmult}        & 33 & 3 (3) & 1.5  & \yes  & 78 \\ 
			\texttt{mergeinter}        & 23 & 2 (1) & 3.4  & \no  & 56 \\ 
			\texttt{mergeintbug}       & 23 & 2 (1) & 2.6  & \no  & 59 \\ 
			\texttt{memcopy}           & 16 & 2 (1) & 1.6  & \yes  & 28 \\ 
			\texttt{multarray}         & 26 & 2 (2) & 2.1  & \yes  & 40 \\ 
			\texttt{parcopy}           & 20 & 2 (1) & 1.2  & \yes  & 30 \\ 
			\texttt{pararray}          & 20 & 2 (1) & 1.2  & \yes  & 31 \\ 
			\texttt{parCopyEven}       & 22 & 2 (1) & 5.0  & \very & 79 \\ 
			\texttt{parMatrix}         & 35 & 4 (2) & 1.1  & \yes  & 80 \\ 
			\texttt{parNested}         & 31 & 4 (2) & 0.5  & \no   & 57 \\ 
			\texttt{relax}             & 33 & 1 (1) & 1.4  & \very & 55 \\ 
			\texttt{reverse}           & 21 & 2 (1) & 3.9  & \yes  & 42 \\ 
			\texttt{reverseBug}        & 21 & 2 (1) & 1.7  & \yes  & 42 \\ 
			\texttt{sanfoundry}        & 27 & 2 (1) & 2.1  & \yes  & 37 \\ 
			\texttt{selectSort}        & 26 & 2 (2) & 1.0 & \no   & 38 \\ 
			\texttt{strCopy}           & 16 & 2 (1) & 0.9  & \no   & 21 \\ 
			\texttt{strLen}            & 10 & 1 (1) & 0.8  & \no   & 15 \\ 
			\texttt{swap}              & 15 & 1 (1) & 1.5  & \yes  & 19 \\ 
			\texttt{swapBug}           & 15 & 1 (1) & 1.5  & \yes  & 19 \\ 
			\bottomrule
		\end{tabular}
	\end{center}
\end{footnotesize}
\caption{Experimental results. For each program, we list the lines of
  code and the number of loops (in brackets the nesting depth).  We
  report the relative size of the inferred specifications compared to
  hand-written specifications, and whether the inferred specifications
  are precise (a star next to the tick indicates slightly more precise
  than hand-written specifications). Inference times are given in ms.}
	\label{tab:evaluation}
\end{table}


We have developed a prototype implementation of our permission inference. The tool is written in Scala and accepts programs written in the Viper language \cite{MuellerSchwerhoffSummers16}, which provides all the features needed for our purposes.

Given a Viper program, the tool first performs a forward numerical
analysis to infer the over-approximate loop invariants needed for our
handling of loops.  The implementation is parametric in the numerical
abstract domain used for the analysis; we currently support the
abstract domains provided by the \textsc{Apron}
library~\cite{JeannetM09}. As we have yet to integrate the
implementation of under-approximate invariants (e.g.,
\cite{Mine12}), we rely on user-provided invariants, or
assume them to be \falseval{} if none are provided.
In a second step, our tool performs the inference and maximum elimination.
Finally, it annotates the input program with the inferred specification.


We evaluated our implementation on 43 programs taken from various sources; included are all programs that do not contain strings from the array memory safety category of SV-COMP 2017, all programs from Dillig et al.~\cite{DilligDA10} (except three examples involving arrays of arrays), loop parallelisation examples from VerCors~\cite{BlomDH15}, and a few programs that we crafted ourselves. We manually checked that our soundness condition
holds for all considered programs. The parallel loop examples were encoded as two consecutive loops where the first one models the forking of one thread per loop iteration (by iteratively exhaling the permissions required for all loop iterations), and the second one models the joining of all these threads (by inhaling the permissions that are left after each loop iteration). For the numerical analysis we used the \emph{polyhedra abstract domain} provided by \textsc{Apron}. The experiments were performed on a dual core machine with a 2.60 GHz Intel Core i7-6600U CPU, running Ubuntu~16.04.

An overview of the results is given in~\tabref{evaluation}. For
each program, we compared the size and precision of the inferred
specification with respect to hand-written ones. The running times
were measured by first running the analysis 50 times to warm up the
JVM and then computing the average time needed over the next 100
runs. The results show that the inference is very efficient.
The inferred
specifications are concise for the vast majority of the
examples. In 35 out of 48 cases, our inference
inferred precise specifications. Most of the imprecisions are due to
the inferred numerical loop invariants. In all cases, manually
strengthening the invariants yields a precise specification. In one
example, the source of imprecision is our abstraction of array-dependent
conditions (see \secref{loops}).


\section{Related Work}
\label{sec:related}

Much work is dedicated to the analysis of
array programs, but most of it focuses on array content, whereas
we infer permission specifications. The
simplest approach consists of ``smashing'' all array elements into a
single memory location \cite{BertraneCCFMMR10}. This is generally
quite imprecise, as only weak updates can be performed on the smashed
array. A simple alternative is to consider array elements as distinct
variables \cite{BertraneCCFMMR10}, which is feasible only when the
length of the array is statically-known. More-advanced approaches perform
syntax-based \cite{GopanRS05,HalbwachsP08,JhalaM07} or
semantics-based \cite{CousotCL11,LiuR17} partitions of an array into
symbolic segments. These require segments to be contiguous (with the
exception of \cite{LiuR17}), and do not easily generalise to
multidimensional arrays, unlike our approach.
Gulwani et al.\ \cite{GulwaniMT08} propose an approach for inferring quantified invariants for arrays by lifting quantifier-free abstract domains. Their technique requires templates for the invariants.

Dillig et al.\ \cite{DilligDA10} avoid an explicit array partitioning by maintaining constraints that over- and under-approximate the array elements being updated by a program statement. Their work employs a technique for directly generalising the analysis of a single loop iteration (based on quantifier elimination), which works well when different loop iterations write to disjoint array locations. Gedell and H\"ahnle \cite{GedellH06} provide an analysis which uses a similar criterion to determine that it is safe to parallelise a loop, and treat its heap updates as one bulk effect. The condition for our projection over loop iterations is weaker, since it allows the same array location to be updated in multiple loop iterations (like for example in sorting algorithms). Blom et al.\ \cite{BlomDH15} provide a specification technique for a variety of parallel loop constructs; our work can infer the specifications which their technique requires to be provided.
%
%

Another alternative for generalising the effect of a loop iteration is to use a first order theorem prover as proposed by Kov\'acs and Voronkov \cite{KovacsV09}. In their work, however, they did not consider nested loops or multidimensional arrays. Other works rely on loop acceleration techniques \cite{AlbertiGS13,BozgaHIKV09}. In particular, like ours, the work of Bozga et al.\ \cite{BozgaHIKV09} does not synthesise loop invariants; they directly infer post-conditions of loops with respect to given preconditions, while we additionally infer the preconditions. The acceleration technique proposed in \cite{AlbertiGS13} is used for the verification of array programs in the tool \textsc{Booster} \cite{AlbertiGS14}.

Monniaux and Gonnord \cite{MonniauxG16} describe an approach for the verification of array programs via a transformation to array-free Horn clauses. Chakraborty et al.\ \cite{ChakrabortyGU17} use heuristics to determine the array accesses performed by a loop iteration and split the verification of an array invariant accordingly. Their non-interference condition between loop iterations is similar to, but stronger than our soundness condition (\cf{}~\secref{loops}).
Neither work is concerned with specification inference.

A wide range of static/shape analyses employ tailored separation logics as abstract domain (\eg{}, \cite{smallfoot,GulavaniCRN09,Calcagno'11,Le'14,Rowe'17}); these works handle recursively-defined data structures such as linked lists and trees, but not random-access data structures such as arrays and matrices. Of these, Gulavani et al.~\cite{GulavaniCRN09} is perhaps closest to our work: they employ an integer-indexed domain for describing recursive data structures. It would be interesting to combine our work with such separation logic shape analyses. The problems of automating biabduction and entailment checking  for array-based separation logics have been recently studied by Brotherston et al.~\cite{Brotherston'17} and Kimura et al.~\cite{KimuraT17}, but have not yet been extended to handle loop-based or recursive programs.

\section{Conclusion and Future Work}
\label{sec:conclusion}
%

We presented a precise and efficient permission inference for array
programs. Although our inferred specifications contain redundancies in
some cases, they are human readable. Our approach integrates well with
permission-based inference for other data structures and with
permission-based program verification.

As future work, we plan to use SMT solving to further simplify
our inferred specifications, to support arrays of
arrays, and to extend our work to an inter-procedural analysis and explore its combination with biabduction techniques.
%


\subsubsection{Acknowledgements.}
We thank Seraiah Walter for his earlier work on this topic, and Malte Schwerhoff and the anonymous reviewers for their comments and suggestions. This work was supported by the Swiss National Science Foundation.

%
%
%


\newpage 
\bibliographystyle{abbrv}


%
%

\bibliography{oo}
\newpage
\appendix
\section{Auxiliary Inference Definitions}\label{sec:auxinference}

\begin{definition}[Sufficient Permission Preconditions]\label{def:sufficient}
A permission expression $p$ denotes a \emph{sufficient permission
  precondition} for a statement $s$ iff, for all states
\state{H}{P}{`s}\ satisfying $\forall v_a,j. \lookup{P}{v_a}{j} \geq
\evalExp{p}{H}{\upd{\upd{`s}{\a}{v_a}}{\qi}{j}}$, we have:
$\config{s}{\state{H}{P}{`s}} \not\reducesStar \permfail$.
\end{definition}

Here, $p$ may mention the designated variables $\a$ and $\qi$ to denote
the memory location of array $\a$ at index $\qi$, and
$\evalExp{p}{H}{\upd{\upd{`s}{\a}{a}}{\qi}{n}}$ denotes the evaluation
of $p$ in the given heap and environment.

\begin{definition}[Guaranteed Permission Postconditions]\label{def:necessary}
If $p$ is a sufficient permission precondition for a statement $s$ then a
permission expression $p'$ is a corresponding \emph{guaranteed
  permission postcondition} for $s$ w.r.t.\ $p$ iff the following
condition holds: For all initial states
\state{H}{P}{`s}\ satisfying $\forall v_a,j. \lookup{P}{v_a}{j} \geq
\evalExp{p}{H}{\upd{\upd{`s}{\a}{v_a}}{\qi}{j}}$, and for all
final states \state{H'}{P'}{`s'} such that
$\config{s}{\state{H}{P}{`s}}\reducesStar\config{\noop}{\state{H'}{P'}{`s'}}$,
we have: $\forall v_a,j. \lookup{P'}{v_a}{j} \geq
\evalExp{p'}{H}{\upd{\upd{`s}{\a}{v_a}}{\qi}{j}}$.
 \end{definition}

Note that guaranteed permission postconditions are expressed in terms
of pre-states are, thus, are evaluated in $H$ and $`s$ (rather than
$H'$ and $`s'$.

\begin{figure}[!t]
\begin{equation*}
\begin{array}{rcl}
\wpp{\noop}{p} & = & p\\
\wpp{\seq{s_1}{s_2}}{p} & = & \wpp{s_1}{\wpp{s_2}{p}}\\
\wpp{\assign{x}{e}}{p} & = & \sub{p}{e}{x}\\
\wpp{\assign{x}{\deref{a}{e}}}{p} & = & \max(\sub{p}{\deref{a}{e}}{x}, \accperm{a}{e}{\rd})\\
\wpp{\assign{\deref{a}{e}}{x}}{p} & = & \max(\psub{p}{a'}{e'}{\cond{e=e'\wedge a=a'}{x}{\deref{a'}{e'}}}, \accperm{a}{e}{1}) \\
\wpp{\exhale{a,e,p'}}{p} &=& p + \accperm{a}{e}{p'}\\
\wpp{\inhale{a,e,p'}}{p} & = & \max(0, \underhavoc{p}{a}{e} - \accperm{a}{e}{p'}) \\
\wpp{\ifcond{b}{s_1}{s_2}}{p} & = & \cond{b}{\wpp{s_1}{p}}{\wpp{s_2}{p}} \\[0.5em]
\diff{\noop}{p} & = & p \\
\diff{\assign{x}{e}}{p} & = & \sub{p}{e}{x} \\
\diff{\assign{x}{\deref{a}{e}}}{p} & = & \sub{p}{\deref{a}{e}}{x} \\
\diff{\assign{\deref{a}{e}}{x}}{p} & = & \psub{p}{a'}{e'}{\cond{e=e'\wedge a=a'}{x}{\deref{a'}{e'}}} \\
\diff{\exhale{a,e,p'}}{p} & = & p - \accperm{a}{e}{p'} \\
\diff{\inhale{a,e,p'}}{p} & = & \overhavoc{p}{a}{e} + \accperm{a}{e}{p'} \\
\diff{\ifcond{b}{s_1}{s_2}}{p} & = & \cond{b}{\diff{s_1}{p}}{\diff{s_2}{p}}
\end{array}
\end{equation*}
\caption{The full inhale statement rules for the permission preconditions and relative permission differences.}
\label{fig:full-wpp-delta}
\end{figure}

\subsection{Conditional Approximation}\label{sec:restrictionapp}

For the handling of loops we need to abstract away array lookups in order to account for the possibility that the corresponding array value is changed by the loop. Next, we describe the operators used to facilitate that. For every boolean expression $b$, we define an over-approximation $\overapprox{b}$ and and under-approximation $\underapprox{b}$ such that $b \entails \overapprox{b}$ and $\underapprox{b} \entails b$ independently of the program state. The over-approximation of a comparison $e_1 \mathbin{op} e_2$, where $op \in \{=, \neq, <, \leq, >, \geq \}$, is given by
\begin{equation*}
	\overapprox{(e_1 \mathbin{op} e_2)} =
	\begin{cases}
		\trueval & \text{if $e_1$ or $e_2$ contains any array lookup} \\
		e_1 \mathbin{op} e_2 & \text{otherwise.}
	\end{cases}
\end{equation*}
The under-approximation $\underapprox{(e_1 \mathbin{op} e_2)}$ is defined completely analogously with the only difference that the $\trueval$ is replaced with $\falseval$. For all remaining boolean expressions the approximation is defined recursively; for instance  $\overapprox{(b_1 \wedge b_2)} = \overapprox{b_1} \wedge \overapprox{b_2}$ and $\overapprox{(\neg b)} = \neg (\underapprox{b})$.

We extend the notion of over- and under-approximation to permission expressions $p$. Here, we want that $p \leq \overapprox{p}$ and $\underapprox{p} \leq p$ hold independently of the program state. Again, these operators are defined recursively on the structure of the permission expression in a straight forward manner, two of the more complicated cases are $\overapprox{\cond{b}{p_1}{p_2}} = \max(\cond{\overapprox{b}}{\overapprox{p_1}}{0}, \cond{\overapprox{(\neg b)}}{\overapprox{p_2}}{0})$ and $\overapprox{(p_1 - p_2)} = \overapprox{p_1} - \underapprox{p_2}$.

Now, assume that $p$ is the precondition or difference inferred for the code after an $\inhale{a,e,p'}$ statement. If the permission expression $p$ mentions the array value $\deref{a}{e}$ then it might be the case that we do not have the required permissions to talk about $\deref{a}{e}$ in the state before the $\inhale$ statement. To account for this, we introduce two refined abstraction operators that over- and under-approximate all array lookups referring to $\deref{a}{e}$ in a permission expression $p$. These operators, denoted $\overhavoc{p}{a}{e}$ and $\underhavoc{p}{a}{e}$, are defined similarly to $\overapprox{p}$ and $\underapprox{p}$, respectively. The only difference to the regular over- and under-approximation operators are the base cases for comparisons:
\begin{equation*}
	\overhavoc{(e_1 \mathbin{op} e_2)}{a}{e}
	=
	\cond{	\bigvee_{(a',e') \in A} (a'=a \wedge e'=e)}{\trueval}{e_1 \mathbin{op} e_2},
\end{equation*}
where $A$ is the set of all tuples $(a', e')$ such that the array access $\deref{a'}{e'}$ appears in $e_1 \mathbin{op} e_2$. In the expression above, we use $\cond{b_0}{b_1}{b_2}$ as a shorthand for $(\neg b_0 \vee b_1) \wedge (b_0 \vee b_2)$. Again, the under-approximate version $\underhavoc{(e_1 \mathbin{op} e_2)}{a}{e}$ is defined analogously but with $\trueval$ replaced with $\falseval$.

\subsection{Full Permission Inference Definitions}\label{sec:wpp-full}
See \figref{full-wpp-delta} for the full definitions of our permission precondition and relative permission difference inference. The differences with respect to \figref{wpp-delta} are in the \code{inhale} cases, which take account for the case of holding \emph{no} permission to a particular location before the \code{inhale} statement --- if it is only the \code{inhale} operation which makes the location accessible afterwards, then one needs to model that other threads/method invocations could have been modifying this location in the meantime. This is reflected by soundly eliminating any dependency on the corresponding array location, when pushing information backwards over the \code{inhale} statement.

\begin{definition}[Auxiliary Permission Notions]\label{def:permissionmaps}
For any heap $H$, environment $`s$, permission map $P$ and permission-expression $p$, we write $P \more \evalExp{p}{H}{`s}$ to mean, for all array values $v_a$ and integer values $v_i$, $\lookup{P}{v_a}{v_i} \more \evalExp{p}{H}{\upd{\upd{`s}{\a}{v_a}}{\qi}{v_i}}$ (\ie{}, the permission map $P$ holds more permissions, pointwise.\\
For two permission expressions $p_1$ and $p_2$, we write $\evalExp{p_1}{H}{`s} \more \evalExp{p_2}{H}{`s}$ to denote an analogous pointwise inequality.\\
For permission maps $P_1$ and $P_2$, we write $P_1 - P_2$ to denote the permission map defined by the pointwise subtraction of the values in the two maps.
\end{definition}

We introduce a definition of a statement being \emph{well-behaved} with respect to our analysis results: effectively, a well-behaved statement is one for which our analysis is sound (we later show that all statements are well-behaved, in this sense, provided that the hypotheses of \thmref{soundness} hold).
\begin{definition}[Well-behaved statements]\label{def:wellbehaved}
A statement $s$ is \emph{well-behaved} if $\forall p_1, p_2, H, P, `s, H', P', s'$, if $P \more \evalExp{\wpp{s}{p_1}}{H}{`s}$ then:
\begin{enumerate}
\item $\config{s}{\state{H}{P}{`s}} \not\reducesStar \permfail$, and,
\item If $\config{s}{\state{H}{P}{`s}} \reducesStar \config{s'}{\state{H'}{P'}{`s'}}$, then:
\begin{enumerate}
\item $P' \more \evalExp{\wpp{s'}{p_1}}{H'}{`s'}$, and,
\item $(P' - P) \more (\evalExp{\diff{s}{p_2}}{H}{`s} - \evalExp{\diff{s'}{p_2}}{H'}{`s'}$
\end{enumerate}
\end{enumerate}
\end{definition}

The following lemma reflects the definition of our $\wpp$ and $\diff$ operators for sequential compositions:
\begin{lemma}[Composition of well-behaved statements]
For any well-behaved statements $s_1,s_2,\ldots,s_n$ (and any $n > 0$), the iterated sequential composition $\seq{s_1}{\seq{s_2}{\seq{\ldots}{s_n}}}$ is well-behaved.
\begin{proof}
By straightforward induction on $n$.
\end{proof}
\end{lemma}

\setcounter{theorem}{0}
\begin{theorem}
Let $\while{b}{s}$ be an exhale-free loop, let
$\Vec{x}$ be the integer variables modified by $s$, and let
$\Invplus$ be a sound over-approximating numerical loop invariant
(over the integer variables in $s$). Then $\inlinemaxunder{\Vec{x}
  \mid \Invplus \wedge b}{\restrict{\wpp{s}{0}}}$ is a sufficient
permission precondition for $\while{b}{s}$.
\begin{proof}
We show here the (technically weaker) result, that \emph{if} $\wpp{s}{0}$ is a sufficient permission precondition for $s$, then our theorem here holds. This additional hypothesis is obtained from the overall inductive soundness proof of our analysis (\thmref{soundness}, below); from this, the proof of the originally stated theorem directly follows.

We can then prove the result by showing that for any finite number $n$ of unrollings of the loop performed by the execution of $\while{b}{s}$, we do not encounter a permission failure, and retain at least $\inlinemaxunder{\Vec{x}
  \mid \Invplus \wedge b}{\restrict{\wpp{s}{0}}}$ permissions. The proof proceeds by induction on $n$. In the inductive case, since $\Invplus$ is a sound over-approximating loop invariant, we must have $`s\entails \Invplus \wedge b$ (note that these expressions are independent of the heap). Therefore, $\evalExp{\inlinemaxunder{\Vec{x}
  \mid \Invplus \wedge b}{\restrict{\wpp{s}{0}}}}{H}{`s} \more \evalExp{\restrict{\wpp{s}{0}}}{H}{`s} \more \evalExp{\wpp{s}{0}}{H}{`s}$, and so the evaluation of $s$ cannot result in a permission failure, since our hypothesis was that $\wpp{s}{0}$ is a sufficient permission precondition for $s$. Furthermore, since $s$ is exhale-free, we are guaranteed to hold at least as many permissions after execution of $s$ as before, and so we can apply our induction hypothesis to conclude.
  \end{proof}
\end{theorem}

\begin{lemma}\label{lem:supercool}
Given a loop $\while{b}{s}$, let $\Vec{x}$ be the integer variables
modified in $s$ and let $\Vec{v}$ and $\Vec{v'}$ be two fresh sets of
variables, one for each of $\Vec{x}$. Suppose that the following implication is valid:
\[
\begin{array}{c}
\vecsub{(\Inv\wedge b)}{{v}}{{x}}\ \wedge\ \vecsub{(\Inv\wedge b)}{{v'}}{{x}}\ \wedge\ (\bigvee{\Vec{v\neq v'}})\ \Rightarrow\\
 \max(\vecsub{\restrict{\wpp{s,0}}}{{v}}{{x}}, \vecsub{\restrict{\wpp{s,0}}}{{v'}}{{x}}) \geq \vecsub{\restrict{\wpp{s}{\vecsub{\restrict{\wpp{s}{0}}}{{v'}}{{x}}}}}{{v}}{{x}}\end{array}
\]
Then, it follows that, for any $n \geq 0$, and $n$ fresh sets of variables $\Vec{v_1},\Vec{v_2},\ldots,\Vec{v_n}$, the following implication is also valid:
\[
\begin{array}{c}
\bigwedge_{1\leq j \leq n}{\vecsub{(\Inv\wedge b)}{{v_j}}{{x}}}\ \wedge\ (\bigwedge_{1\leq j < k \leq n}{\bigvee{\Vec{v_j\neq v_k}}})\ \Rightarrow\\
 (\maxunder{1\leq j \leq n}{\vecsub{\restrict{\wpp{s,0}}}{{v_j}}{{x}}}) \geq \vecsub{\restrict{\wpp{s}{\vecsub{\restrict{\wpp{s}{\ldots\vecsub{\restrict{\wpp{s}{0}}}{{v_n}}{{x}}\ldots}}}{{v_2}}{{x}}}}}{{v_1}}{{x}}\end{array}
\]
\begin{proof}
The proof follows by straightforward induction on $n$.
\end{proof}
\end{lemma}

\begin{theorem}[Soundness Condition for Loop Preconditions]
Given a loop $\while{b}{s}$, let $\Vec{x}$ be the integer variables
modified in $s$ and let $\Vec{v}$ and $\Vec{v'}$ be two fresh sets of
variables, one for each of $\Vec{x}$. Then $\inlinemaxunder{\Vec{x}
  \mid \Invplus \wedge b}{\restrict{\wpp{s}{0}}}$ is a sufficient
permission precondition for $\while{b}{s}$ if the following implication
is valid in all states:
\[
\begin{array}{c}
\vecsub{(\Inv\wedge b)}{{v}}{{x}}\ \wedge\ \vecsub{(\Inv\wedge b)}{{v'}}{{x}}\ \wedge\ (\bigvee{\Vec{v\neq v'}})\ \Rightarrow\\
 \max(\vecsub{\restrict{\wpp{s,0}}}{{v}}{{x}}, \vecsub{\restrict{\wpp{s,0}}}{{v'}}{{x}}) \geq \vecsub{\restrict{\wpp{s}{\vecsub{\restrict{\wpp{s}{0}}}{{v'}}{{x}}}}}{{v}}{{x}}\end{array}
\]
\begin{proof}
As for the previous theorem, we show here the weaker result (which nonetheless fits into our overall inductive soundness argument), that \emph{if}, for any $n$ sequential compositions of $s$ with itself $\seq{s}{\seq{s}{\seq{\ldots}{s}}}$, the permission expression $\wpp{\seq{s}{\seq{s}{\seq{\ldots}{s}}}}{0}$ is a sufficient permission precondition for $\seq{s}{\seq{s}{\seq{\ldots}{s}}}$, then the desired result holds (the proof then essentially shows that our precondition for the whole loop is at least as large as the precondition for this unrolled statement). The theorem in its originally-stated form is again a corollary of $\thmref{soundness}$.

We show that, for any such $n$ unrollings of the loop, $\entails \inlinemaxunder{\Vec{x}
  \mid \Invplus \wedge b}{\restrict{\wpp{s}{0}}} \geq \wpp{\seq{s}{\seq{s}{\seq{\ldots}{s}}}}{0}$. Since each initial state of a loop execution is guaranteed to satisfy $\Invplus \wedge b$, and without loss of generality (adding a loop counter if necessary, as discussed in \secref{loops}), we can assume that these states are pairwise distinct when considering only the values of variables, we obtain the result from \lemref{supercool}, above.
\end{proof}
\end{theorem}

Our overall soundness result is ultimately handled by the following technical lemma:
\begin{lemma}\label{lem:amazing}
For all statements $s$, if all loops occurrings in $s$ are either exhale-free or satisfy the condition of \thmref{condition}, then $s$ is well-behaved.
\begin{proof}
In order to prove the result, we need to instrument the program in question with auxiliary information: we annotate each loop with a pair $(\Invplus,\Invminus)$, which are initially its over- and under-approximate loop invariants, respectively. We need to record both the \emph{original} forms of these invariants, and a pair of invariants which we use as auxiliary state during execution of the corresponding loop. This \emph{updateable} pair of loop invariants is adjusted during execution: every time a new iteration of the loop is started, we conjoin an additional conjunct to both invariants, expressing that this particular initial state of the loop iteration cannot be seen again. The rules for $\wpp$ and $\diff$ for the remaining loop iterations take account of the currently-updated versions of these two invariants; this is critical for the inductive step to go through. For example, this allows a loop iteration to exhale some permissions, and for the remaining copy of the loop code not to require the permissions for exactly that iteration again. Since this instrumentation does not affect the steps of program execution, we can be sure that our result also holds for the uninstrumented version of the operational semantics.

The proof then proceeds by structural induction on $s$ (see also the comment below the lemma). The parameterisation of the definition of well-behaved statements by the additional permission expressions $p_1$ and $p_2$ allows for the inductive cases for sequential compositions to be handled simply: we can plug in the appropriate intermediate results computed by our $\wpp$ and $\diff$ operators. The case for loops is (as expected), the most involved, and uses either \thmref{stable} or \thmref{condition} to rule out the possibility of permission failures. The fact that the gain or loss of permissions during execution results in a state with at least as many permissions as predicted by our $\diff$ operator follows from the inductive argument.
\end{proof}
\end{lemma}
It may be surprising that the proof runs by structural induction on $s$, even in a language with loops; the reason this works is that the notion of being well-behaved builds in all finite unrollings of the statement, and our lemmas and case regarding projecting our specifications for loops need to handle all such unrollings at once, in the proof.

Our main soundness result is now a direct corollary of the above lemma:
\begin{theorem}[Soundness of Permission Inference]
For any statement $s$, if every $\whilesym$ loop in $s$ either is
exhale-free or satisfies the condition of \thmref{condition} then
$\wpp{s}{0}$ is a sufficient permission precondition for $s$, and
$\wpp{s}{0} + \diff{s}{0}$ is a corresponding guaranteed permission
postcondition.
\begin{proof}
By \lemref{amazing}, $s$ is well-behaved; instantiating both $p_1$ and $p_2$ in the definition of being well-behaved yields the result directly.
\end{proof}
\end{theorem}

\section{Auxiliary Maximum Elimination Definitions and Proofs}\label{sec:maxelimextras}\label{sec:qebackground}

\subsection{Simplified Grammars}\label{sec:simplified}

The first step in our solution is to reduce a general maximum-elimination problem to several smaller maximum problems over simpler grammars of permission and boolean expressions; these simplified grammars are defined as follows:
\begin{definition}[Simple boolean and permission expressions over $x$.]\label{def:simple}
The grammars of \emph{simple boolean expressions over $x$} and \emph{simple permission expressions over $x$}, are defined as follows:
\[
\begin{array}{rcl}
\simple{b}{x} &\defequals& \simple{b_1}{x} \wedge \simple{b_2}{x} \mid \simple{b_1}{x} \vee \simple{b_2}{x} \mid \divides{n}{(x {+} e)} \mid \notdivides{n}{(x {+} e)} \mid x\;\op\; e \mid b'\\
&\multicolumn{2}{l}{\textit{where }x\not\in\FV{b'}\textit{ and, in all cases }x\notin\FV{e}\quad\quad\op\in\{=, \neq, <, \leq, >, \geq\}}\\
\simple{p}{x} &\defequals& \cond{\simple{b}{x}}{r}{0} \mid \simple{p_1}{x} + \simple{p_2}{x} \mid \simple{p_1}{x} - \cond{\simple{b}{x}}{r}{0} \mid \\
&&\max(\simple{p_1},\simple{p_2}) \mid \min(\simple{p_1},\simple{p_2})\quad\quad\quad r\;\defequals\;q \mid \rd\\
\end{array}\]
Simple permission expressions must additionally satisfy the restriction that no addition operators have subexpressions with subtraction operators inside them.

\noindent
A maximum expression $\inlinemaxunder{x\suchthat b}{p}$ is \emph{simple} if both $b$ and $p$ are simple over $x$.

\noindent
 We write $\trueexp$ and $\falseexp$ as shorthands for $0=0$ and $0\neq 0$, respectively. For simple boolean expressions $\simple{b}{x}$, we write $\neg \simple{b}{x}$ as shorthand for the simple boolean expression obtained by pushing the negation down, flipping connectives and literals to their duals, \eg{} $\neg (\divides{n}{(x + e)} \wedge x > 0)$ is shorthand for $\notdivides{n}{(x+e)}\vee x \leq 0$.
\end{definition}

The reduction of boolean expressions to simple boolean expressions is handled following the initial steps of Cooper's Quantifier Elimination procedure \cite{CooperQE}, briefly summarised as follows. Firstly, each boolean expression should be converted to negation normal form (pushing all negations down to individual literals). Next, all occurrences of $x$ in literals are grouped such that there is at most one occurrence (\eg{} rewriting $x + 3 > 2 - x$ as $2\cdot x > -1$). If the resulting boolean expressions contain different coefficients for $x$, these literals are then ``multiplied out'' such that each contains only $d\cdot x$ for some common integer constant $d$. If $d\neq 1$, we then replace $d\cdot x$ with a fresh variable $x'$ throughout, and conjoin $\divides{d}{x'}$ to the condition $b'$ filtering the maximum expression. For each converted boolean expression, the values of $x'$ making the resulting expression true are exactly the values $x'/d$ of $x$ making the original boolean expression true. This multiplying-out step needs to be performed across the entire maximum expression in question; if several such steps are performed, then the multiplication is by the lowest common multiple of them all.

%
We reduce the problem of finding a method for eliminating general ${\inlinemaxunder{x\suchthat b}{p}}$ expressions to that of eliminating \emph{simple} maximum expressions, as follows. For the permission expression $p$ in question, we first distribute any $+$ and $-$ operators over $\max$, $\min$ and conditional expressions in which the second branch is not simply $0$. For example, we rewrite $\max(p_1,p_2) + p_3$ as $\max(p_1+p_3,p_2+p_3)$. In this way, we obtain an expression ${\inlinemaxunder{x\suchthat b}{p}}$ in which any occurrences of $\max$, $\min$ and conditional expressions of this are nested at the top level. Note that propagating subtraction over a $\max$ as its second operand requires the introduction of a $\min$ term - this is the motivation for including them in our grammar. The restriction of \defref{simple} that subtraction operators do not occur under addition operators can be achieved with further rewritings; \eg{} rewriting $p_1 + (p_2 - \leaf{b}{r})$ to $(p_2 + p_1) - \leaf{b}{r}$.

 We now eliminate conditional expressions, and as many $\max$ operators as possible, by repeatedly applying the following two equalities (read from left-to-right) throughout the permission expression:
\[
\begin{array}{rcl}
\maxunder{x\suchthat b}{\max(p_1,p_2)} &=& \max(\maxunder{x\suchthat b}{p_1}, \maxunder{x \suchthat b}{p_2})\\
\maxunder{x\suchthat b}{\cond{b'}{p_1}{p_2}} &=& \left\{\begin{array}{ll}
\cond{b'}{\maxunder{x\suchthat b}{p_1}}{\maxunder{x\suchthat b}{p_2}} & \textit{ if }p_2\neq 0\textit{ and }x\notin\FV{b'}\\
\max(\maxunder{x\suchthat b\wedge b'}{p_1}, \maxunder{x\suchthat b\wedge \neg b'}{p_2}) &  \textit{ if }p_2\neq 0\textit{ and }x\in\FV{b'}\\\end{array}\right.
\end{array}
\]
This yields a number of pointwise maximum terms to solve, each of which is simple over $x$. For example, when computing a precondition for the loop in the \code{copyEven} example (projecting the precondition for the loop body using the loop condition and integer invariant), we generate the pointwise maximum $\maxunder{\code{j} \suchthat 0\leq \code{j} \wedge \code{j} < \length{\code{a}}}{\cond{\divides{2}{\code{j}}}{\cond{\a=\code{a}\wedge\qi=\code{j}}{\rd}{0}}{\cond{\a=\code{a}\wedge\qi=\code{j}}{1}{0}}}$ which we rewrite here to:
$\begin{array}{ll}\max(&\maxunder{\code{j} \suchthat 0\leq \code{j} \wedge \code{j} < \length{\code{a}} \wedge \divides{2}{\code{j}}}{\cond{\a=\code{a}\wedge\qi=\code{j}}{\rd}{0}},\\
&\maxunder{\code{j} \suchthat 0\leq \code{j} \wedge \code{j} < \length{\code{a}} \wedge \notdivides{2}{\code{j}}}{\cond{\a=\code{a}\wedge\qi=\code{j}}{1}{0}})\end{array}$

\subsection{Quantifier Elimination and Maximum Elimination Lemmas}

\begin{figure}[t]
\[
\begin{array}{rcccl}
\multicolumn{5}{c}{\mini{b} = (\emptyset,1)\textit{ if }x\notin\FV{b}}\\
\mini{\x=e} &=& \mini{\x \geq e} &=& (\{e\},1)\\
\mini{\x\neq e} &=& \mini{\x > e} &=& (\{e{+}1\},1)\\
\mini{\x \leq e} &=& \mini{\x < e} &=& (\emptyset,1)\\
\mini{\divides{n}{(x + e)}} &=& \mini{\notdivides{n}{(x+e)}} &=& (\emptyset,n)\\
\mini{b_1\wedge b_2} &=& \mini{b_1\vee b_2} &=& (S_1\union S_2, \lcm{`d_1,`d_2})\\
\multicolumn{5}{l}{\quad\quad\quad\textit{where}\quad
(S_1,`d_1) = \mini{b_1}\quad\textit{and}\quad
(S_2,`d_2) = \mini{b_2}}\\
\end{array}
\]
\caption{Boundary Expression Selection for Simple Boolean Expressions}\label{fig:mini}
\end{figure}

\begin{definition}[Boundary Expression Selection for Booleans \cite{CooperQE}]\label{def:miniboolean}
The \emph{boundary expression selection for $x$ in $\simple{b}{x}$} is written $\mini{\simple{b}{x}}$, returning a pair of a set $S$ of integer-typed expressions, and an integer constant $`d$, as defined in \figref{mini}.
\end{definition}
For \eg{} the literal $x=4$, it is clear that the value $4$ is the smallest value for $x$ making the literal true. For $x\neq 4$, the boundary value $5$ is generated; although there are smaller values making the literal true, this is the only one with the property that the next smallest value does \emph{not}. To understand the role of the $`d$ result returned, consider the input expression $x\neq 4 \wedge \divides{3}{x}$. While $5$ remains a boundary value for the first literal, this value does not make the whole formula true; it ``misses'' the divisibility constraint by being slightly too small. For this reason, the actual set of terms represented by the output pair $(S,`d)$ should be understood to be $\{e + d \mid e\in S \wedge d \in \range{0}{`d-1}\}$; the $`d$ result is used to catch these missing cases. For the above formula, the resulting set would be $\{5,6,7\}$, which indeed includes the smallest value for $x$ satisfying $x\neq 4 \wedge \divides{3}{x}$.

\begin{lemma}\label{lem:notesone}
For all simple boolean expressions $b$, environments $`s$ and integer expressions $n$, if $(S,`d) = \mini{b}$ and $`s\entails \sub{b}{n}{x}$ then, for all integer constants $m > 0$, either:
\begin{enumerate}
\item $\exists e\in S, d\in \range{0}{m\by`d - 1}.\ `s\entails n = e + d$, or:
\item $`s\entails \sub{b}{(n-m\by `d)}{x}$
\end{enumerate}
\begin{proof}
By straightforward induction over the definition of $\mini{b}$ (note that the structure of $b$ decreases on every recursive call (\cf{} \figref{mini}). The $\forall m$ quantification is a necessary generalisation for the inductive argument to go through, to handle the recalculation of $`d$ to be the lowest-common-multiple of the values from the recursive calls, in the conjunction and disjunction cases.
\end{proof}
\end{lemma}

\begin{lemma}\label{lem:mini}
Given a simple boolean expression $b$ (over $x$), and an environment $`s$, if $`s \entails \exists y. \sub{b}{y}{x} \wedge \forall z. z < y \imp \neg\sub{b}{z}{x}$ then $`s \entails (\exists x. b) \Leftrightarrow \veeunder{\substack{e \in S\\d \in \range{0}{`d -1}}}{\sub{b}{e + d}{x}}$, where $(S,`d) = \mini{b}$.
\begin{proof}
The hypothesis $`s \entails \exists y. \sub{b}{y}{x} \wedge \forall z. z < y \imp \neg\sub{b}{z}{x}$ expresses that the existentially-bound value $y$ is the \emph{smallest} value for $x$ such that $b$ holds; in particular, $`s\entails \sub{b}{y}{x}$. Therefore, by \lemref{notesone}, we must have $\exists e\in S, d\in \range{0}{m\by`d - 1}.\ `s\entails y = e + d$ (since the second case of the lemma contradicts our hypothesis). Therefore, $`s\entails \sub{b}{e+d}{x}$ for some such $e$ and $d$, and the required disjunction is implied.
\end{proof}
\end{lemma}

\begin{definition}[Left-Infinite Projection for Boolean Expressions \cite{CooperQE}]\label{def:lipboolean}
The \emph{left-infinite projection} of $\simple{b}{x}$, written $\leftInf{\simple{b}{x}}$ is defined by the rules of \figref{leftInf}.
\end{definition}

\begin{figure}[t]
\[
\begin{array}{rcccccl}
\multicolumn{7}{c}{\leftInf{b} = (b,1)\textit{ if }x\notin\FV{b}}\\
\leftInf{\x=e} &=& \leftInf{\x \geq e} &=& \leftInf{\x > e} &=& (\falseExp,1)\\
\leftInf{\x\neq e} &=& \leftInf{\x \leq e} &=& \leftInf{\x < e} &=& (\trueExp,1)\\
\leftInf{\divides{n}{(x + e)}} &=& \multicolumn{1}{l}{(\divides{n}{(x + e)},n)} && \multicolumn{1}{r}{\leftInf{\notdivides{n}{(x+e)}}} &=& (\notdivides{n}{(x+e)},n)\\
\leftInf{b_1\wedge b_2} &=& \multicolumn{1}{l}{(b'_1 \wedge b'_2, \lcm{`d_1,`d_2}) } && \multicolumn{1}{r}{\leftInf{b_1\vee b_2}} &=& (b'_1 \vee b'_2, \lcm{`d_1,`d_2})\\
\textit{ where } (b'_1,`d_1) &=& \multicolumn{1}{l}{\leftInf{b_1}} && \multicolumn{1}{r}{\textit{ where } (b'_1,`d_1)} &=& \leftInf{b_1}\\
 (b'_2,`d_2) &=& \multicolumn{1}{l}{\leftInf{b_2}} && \multicolumn{1}{r}{ (b'_2,`d_2)} &=& \leftInf{b_2}\\
\end{array}
\]
\caption{Left Infinite Projection for Simple Boolean Expressions}\label{fig:leftInf}
\end{figure}

\begin{lemma}\label{lem:leftInfBool}
For all simple boolean expressions $b$, and environments $`s$, there exists an integer constant $m$ such that, for all expressions $e$:
\[
`s \entails e\leq m \Rightarrow (\sub{b}{e}{x} \Leftrightarrow \sub{b'}{(e\modSym `d)}{x})
\]
where $(b',`d) = \leftInf{b}$
\begin{proof}
By straightforward induction over the definition of $\leftInf{b}$ (the structure of $b$ is reduced in every recursive call, so this is well-founded).
\end{proof}
\end{lemma}

\begin{lemma}
For a given state $`s$, if $`s \entails \forall y. \sub{b}{y}{x} \imp \exists z. z < y \wedge \sub{b}{z}{x}$ then $`s \entails (\exists x. b) \Leftrightarrow \veeunder{d \in \range{0}{`d -1}}{\sub{b'}{d}{x}}$, where $(b',`d) = \leftInf{b}$.
\begin{proof}
The hypothesis expresses that $b$ can be made true for arbitrarily small values $z$. By \lemref{leftInfBool}, we know that for values $e$ smaller than some $m$,
 evaluating $\sub{b}{e}{x}$ gives the same answer as evaluating $\sub{b'}{(e\modSym `d)}{x}$. Therefore, one of the disjuncts in the statement of this result must hold (since we enumerate all possible values of $(e\modSym `d)$).
\end{proof}
\end{lemma}

\subsection{Left-Infinite Projection of Permission Expressions}\label{sec:leftinf}
\begin{figure}[t]
\[
\begin{array}{rcl}
\leftInf{b} &=& b \textit{ if }x\notin\FV{b}\\
\leftInf{\leaf{b}{r}} &=& (\leaf{b'}{r},`d)\textit{ where }(b',`d) = \leftInf{b}\\
\leftInf{p_1 + p_2} &=& (p_1' + p_2', `d)\textit{ where }\\
&\multicolumn{2}{l}{(p_1',`d_1) = \leftInf{p_1},~(p_2',`d_2) = \leftInf{p_2},~`d=\lcm{`d_1,`d_2}}\\
\leftInf{p - \leaf{b}{r}} &=& (p' - \leaf{b'}{r}, `d)\textit{ where }\\
&\multicolumn{2}{l}{(p',`d_1) = \leftInf{p},~(b',`d_2) = \leftInf{b},~`d=\lcm{`d_1,`d_2}}\\
\leftInf{\max(p_1,p_2)} &=& (\max(p_1',p_2'), `d)\textit{ where }\\
&\multicolumn{2}{l}{(p_1',`d_1) = \leftInf{p_1},~(p_2',`d_2) = \leftInf{p_2},~`d=\lcm{`d_1,`d_2}}\\
\leftInf{\min(p_1,p_2)} &=& (\min(p_1',p_2'), `d)\textit{ where }\\
&\multicolumn{2}{l}{(p_1',`d_1) = \leftInf{p_1},~(p_2',`d_2) = \leftInf{p_2},~`d=\lcm{`d_1,`d_2}}
\end{array}
\]
\caption{Left Infinite Projection for Simple Permission Expressions}\label{fig:leftInfPerm}
\end{figure}
We deal first with the special case of there being \emph{arbitrarily small} values of $x$ \emph{defining} the pointwise maximum expression in question. We define a function to reflect the value of the permission expression in question for these arbitrarily small values. The idea behind the function is that for the supported grammar of expressions, if one considers sufficiently small values of $x$, all evaluations of $p$ will yield the same value, modulo the divisibility constraints in $p$. Our function returns a simplified permission expression (representing the value that $p$ takes for these sufficiently-small values of $x$), along with the lowest common multiple of these divisibility constraints (defining the period of the repeating values of $p$ in this small-enough range):
\begin{definition}[Left-Infinite Projection for Permission Expressions]\label{def:lip}
The \emph{left-infinite projection} of $\simple{p}{x}$, written $\leftInf{\simple{p}{x}}$ returns a pair $(p',`d)$ of permission expression and integer value, and is defined in \figref{leftInfPerm}.
\end{definition}

As an example, suppose we have the $\simple{p}{j}$ permission expression $\leaf{\a=\code{a}\wedge\qi=\code{j}}{\rd}$. We have $\leftInf{\code{j}}{\a=\code{a}\wedge\qi=\code{j}} = (\a=\code{a}\wedge\falseexp,1)$ ($\falseexp$ since the expression $\qi=\code{j}$ is not true for arbitrarily small values of $\code{j}$, and $1$ since there are no divisibility constraints). Simplifying the resulting expression, we obtain $\leftInf{\simple{p}{j}} = (0,1)$. This $0$ result reflects the fact (common for expressions generated by our permission analysis) that the value of the maximum expression of interest is \emph{not} defined by arbitrarily small values of \code{j} (as can be easily seen by hand, the maximum value is defined by choosing $\code{j}$ to be exactly $\qi$.

The following lemma shows how $p$ and its left-infinite projection are related, for sufficiently small values of $x$ (these are values for which the evaluation of all (in/dis-)equalities have already stabilised).

\begin{lemma}\label{lem:leftInfPerm}
For all simple permission expressions $p$ and environments $`s$, there exists an integer constant $m$ such that, for all expressions $e$:
\[
`s \entails e\leq m \Rightarrow (\sub{p}{e}{x} = \sub{p'}{(e\modSym `d)}{x})
\]
where $(p',`d) = \leftInf{p}$
\begin{proof}
By straightforward induction over the definition of $\leftInf{p}$ (the structure of $p$ is reduced in every recursive call, so this is well-founded), using the corresponding \lemref{leftInfBool}.
\end{proof}
\end{lemma}

\begin{lemma}
For a given state $`s$, if $`s \entails \forall y. \sub{b}{y}{x} \imp \exists z. z < y \wedge \sub{b}{z}{x}$ then $`s \entails (\exists x. b) \Leftrightarrow \veeunder{d \in \range{0}{`d -1}}{\sub{b'}{d}{x}}$, where $(b',`d) = \leftInf{b}$.
\begin{proof}
The hypothesis expresses that $b$ can be made true for arbitrarily small values $z$. By \lemref{leftInfBool}, we know that for values $e$ smaller than some $m$,
 evaluating $\sub{b}{e}{x}$ gives the same answer as evaluating $\sub{b'}{(e\modSym `d)}{x}$. Therefore, one of the disjuncts in the statement of this result must hold (since we enumerate all possible values of $(e\modSym `d)$).
\end{proof}
\end{lemma}

The following result states that, when arbitrarily small values of $x$ define $\maxunder{x\suchthat b}{p}$, these can be captured using our left-infinite projection, along with Cooper's analogous definition for boolean expressions:
\begin{lemma}\label{lem:lip}
For any $\simple{p}$, $\simple{b}$, $`s$, if $(p',`d_1) = \leftInf{\simple{p}}$, $(b',`d_2) = \leftInf{\simple{b}}$, and
$`s \entails \forall y. \exists z. z < y \wedge \sub{\simple{p}}{z}{x} > 0 \wedge \sub{\simple{b}}{z}{x} \wedge \sub{\simple{p}}{z}{x} = \maxunder{x \suchthat \simple{b}}{\simple{p}}$, then:\\$\maxunder{x \suchthat \simple{b}}{\simple{p}} = \maxunder{d \in \range{0}{\lcm{`d_1,`d_2}-1}}{\cond{\sub{b'}{d}{x}}{\sub{p'}{d}{x}}{0}}$
\begin{proof}
The (rather long) last hypothesis expresses that there are arbitrarily small values of $x$ defining the desired maximum value. By \lemref{leftInfPerm}, we know that for values $e$ smaller than some $m$, evaluating $\sub{p}{e}{x}$ gives the same answer as evaluating $\sub{p'}{(e\modSym `d_1)}{x}$. Similarly, by \lemref{leftInfBool}, we know that for $e$ smaller than some $m'$, evaluating $\sub{b}{e}{x}$ gives the same answer as evaluating $\sub{b'}{(e\modSym `d_2)}{x}$. Therefore, for all $e$ smaller than the minimum of $m$ and $m'$, the only values $\sub{\cond{b}{p}{0}}{e}{x}$ takes are included within the set $\{\cond{\sub{b'}{d}{x}}{\sub{p'}{d}{x}}{0} \mid d \in \lcm{`d_1,`d_2} - 1 \}$. Since we are working under the hypothesis that arbitrarily small values of $x$ define the desired maximum value, it must also be one of the values in this set. By taking the maximum of all of these, as in the statement of the lemma, we are guaranteed to capture it.
\end{proof}
\end{lemma}

\subsection{Boundary Expressions for Permission Expressions}
We now turn to the results concerning our boundary expression selection for permission expressions (\defref{mini}).

\begin{lemma}\label{lem:notestwo}
For all simple permission expressions $p$, environments $`s$ and integer expressions $n$, if $p$ contains \emph{no subtraction terms}, and $(T,`d) = \mini{p}$ then, for all integer constants $m > 0$, either:
\begin{enumerate}
\item $\exists (e,b)\in T, d\in \range{0}{m\by`d - 1}.\ `s\entails \sub{b}{n}{x} \wedge n = e + d$, or:
\item $`s\entails \sub{p}{(n-m\by `d)}{x} \geq \sub{p}{n}{x}$
\end{enumerate}
\begin{proof}
By straightforward induction on $p$, using \lemref{notesone} when our definition makes use of the analogous notion of boundary expressions for booleans (\defref{miniboolean}). The proof uses as additional lemma that, since $p$ contains no subtraction terms, it (and all of its subexpressions) denote only non-negative values (in all environments), which is used in the inductive case for minimum expressions.
\end{proof}
\end{lemma}

We now prove a slight generalisation of the above result, to account for subtraction terms (in the limited positions that they are allowed; recall that they cannot occur under addition operators, according to \defref{simple}).

\begin{lemma}\label{lem:notesthree}
For all simple permission expressions $p$, environments $`s$ and integer expressions $n$, and $(T,`d) = \mini{p}$ and $`s\entails \sub{p}{n}{x} \geq 0$ then, for all integer constants $m > 0$, at least one of the following holds:
\begin{enumerate}
\item $\exists (e,b)\in T, d\in \range{0}{m\by`d - 1}.\ `s\entails \sub{b}{n}{x} \wedge n = e + d$, or:
\item $`s\entails \sub{p}{(n-m\by `d)}{x} \geq \sub{p}{n}{x}$, or:
\item $`s\entails \sub{p}{n}{x} = 0$.
\end{enumerate}
\begin{proof}
By straightforward induction on $p$, using \lemref{notesone} and \lemref{notestwo} (and considering the combinations of case-splits that these result in).
\end{proof}
\end{lemma}

\setcounter{theorem}{3}
We are now finally in a position to prove \thmref{maxelim}, whose definition we recall here:
\begin{theorem}[Simple Maximum Expression Elimination]
For any pair $(\simple{p}{x},\simple{b}{x})$, if $\entails p \geq 0$, then we have:
\begin{equation*}
\begin{split}
\entails \maxunder{x \suchthat b}{p} = \max\bigl(\maxunder{\substack{(e,b'') \in T \\ d \in \range{0}{\delta-1}}}{\cond{b'' \wedge \sub{b}{e + d}{x}}{\sub{p}{e + d}{x}}{\zeroPerm}}), \\[-10pt]
\maxunder{d \in \range{0}{\lcm{`d_1,`d_2}-1}}{\cond{\sub{b'}{d}{x}}{\sub{p'}{d}{x}}{\zeroPerm}} \bigr)
\end{split}
\end{equation*}
where $(T,`d) = \minimax{(p,b)}$, $(b',`d_1) = \leftInf{b}$ and $(p',`d_2) = \leftInf{p}$.
\begin{proof}
We consider the values of the two equated terms in an arbitrary environment $`s$ (on which some of the following case-splits may depend, due to other free variables in the expressions).

We then consider the same top-level case-split as informs the design of our algorithm: (1) there is a \emph{smallest} value of $x$ defining the pointwise maximum in question, or (2) the pointwise maximum's value is defined by arbitrarily small values of $x$. Note that in either case, it is then sufficient to show that the result of our maximum elimination is \emph{at least as large} as the pointwise maximum in question; since every term in our maximum construction has either a value that $p$ takes for \emph{some} value of $x$, or the value $0$, it certainly cannot result in an answer which is too large.

If $`s\entails \maxunder{x \suchthat b}{p} = 0$, we are immediately done, since our resulting expression has, by construction, a non-negative value in all states. Thus, we proceed with the case that $`s\entails \maxunder{x \suchthat b}{p} > 0$.

Case (2) also follows directly, since \lemref{lip} tells us that the pointwise maximum's value will be equal to $\maxunder{d \in \range{0}{\lcm{`d_1,`d_2}-1}}{\cond{\sub{b'}{d}{x}}{\sub{p'}{d}{x}}{\zeroPerm}}$, which appears as argument to our overall binary $\max$ operator.

Therefore, we are left to consider case (1): there is \emph{some} smallest value of $x$ (say, $n$) such that both $`s\entails \sub{b}{n}{x}$ and $`s\entails \maxunder{x \suchthat b}{p} = \sub{p}{n}{x}$. Note that $`d = \lcm{`d_1,`d_2}$ (our algorithms compute this value two different ways). Let $(T_p,`d_2) = \minimax{p}$ and $(S_b,`d_1) = \mini{b}$ (following \figref{miniPerm}).

We apply \lemref{notesone} (choosing $m$ to be $`d / `d_1$, so that $m\by `d_1 = `d$), to obtain two cases:
either $\exists e\in S_b, d\in\range{0}{`d-1}. `s\entails e + d = n$, or: (II) $`s\entails \sub{b}{(n-`d)}{x}$.

Similarly, from \lemref{notesthree}, we obtain the two cases that either: (I') $\exists (e,b')\in T_p, d\in\range{0}{`d-1}. `s \entails \sub{b}{e}{x} \wedge e + d = n$ or: (II') $`s \entails \sub{p}{(n-`d)}{x} \geq \sub{p}{n}{x}$.

If case (I') holds, we are done, since $T_p\subseteq T$, and so we have captured the defining value for the pointwise maximum in our iteration over the elements of $T$. If cases (II) and (II') both hold, then we contradict our assumption from overall case (1), since the value $n-`d$ is smaller than $n$, satisfies $b$ and results in a larger value for $p$. Therefore, we are left to consider the combination of cases (I) and (II'). By case (I) we have a boundary condition $e$ for $b$, but this boundary condition gets an additional (large disjunction as a) filter condition, according to \figref{miniPerm}; in order to conclude the case, we need to justify that this filter condition is guaranteed to hold (and hence, the boundary condition still gets to contribute to our definition of $T$).

Consider now the set of values $L = \{n-a\by`d \mid a > 0\}$. We identify three cases:
\begin{enumerate}
\item $\exists l\in L. `s\entails \sub{b}{l}{x} \wedge \sub{p}{l}{x} = \sub{p}{n}{x}$: This again contradicts our assumption on $n$.
\item $\forall l\in L. `s\entails \neg\sub{b}{l}{x} \wedge \sub{p}{l}{x} = \sub{p}{n}{x}$: Then by \lemref{leftInfPerm}, there exists $d\in\range{0}{`d-1}$ such that $`s\entails \sub{\leftInf{p}}{d}{x} = \sub{p}{n}{x} \wedge \sub{\neg\leftInf{b}}{d}{x}$, and therefore the first large disjunct in the filter condition for $e$ holds.
\item $\exists l\in L. `s\entails \neg{\sub{b}{l}{x}} \wedge \sub{p}{l}{x} = \sub{p}{n}{x} \wedge \sub{p}{l-`d}{x} < \sub{p}{n}{x}$: Then, by \lemref{notesthree}, $\exists (e'',b'')\in T_p, d''\in\range{0}{`d-1}. `s\entails \sub{b''}{l}{x} \wedge e'' + d'' = l$. Then the second large disjunct in the filter condition for $e$ holds.
\end{enumerate}
In the latter two cases, we therefore have a boundary expression in $T$ whose filter condition holds, and which is guaranteed to define the pointwise maximum in question.
\end{proof}
\end{theorem}

\end{document}